\documentclass[twocolumn]{aastex631}
\usepackage{comment}
\usepackage{alphabeta}
\usepackage{rotating}
\usepackage{tabularx}
\usepackage{tgbonum}
\usepackage{lmodern}
\usepackage{ae,aecompl}
\usepackage{epigraph} 
\usepackage{graphicx}
\usepackage{enumitem}
\usepackage{amsmath,bm}
\usepackage{amssymb}
\usepackage{changepage}
\usepackage{verbatim}
\usepackage{xparse}
\usepackage{float}
\usepackage{physics}
\usepackage[noabbrev]{cleveref}
\usepackage{mathrsfs}
\usepackage{nicefrac}
\usepackage{xspace}
\usepackage{comment}
\usepackage[]{hyperref}
\usepackage{xcolor}

\newcommand{\degree}{\ensuremath{^{\circ}}}
\newcommand{\Ye}{\ensuremath{Y_e}}

\begin{document}

\title{Nucleosynthesis Conditions in Outflows of White Dwarfs Collapsing to Neutron Stars}

\correspondingauthor{Eirini Batziou, H.-Thomas Janka}
\email{batziou@mpa-garching.mpg.de, thj@mpa-garching.mpg.de}

\shorttitle{Outflows of Collapsing White Dwarfs}
\shortauthors{Batziou et al.}

\author[0000-0003-3751-6812]{Eirini Batziou}
\affiliation{Max Planck Institute for Astrophysics, Karl-Schwarzschild-Str.~1, 85748, Garching, Germany}
\affiliation{Technical University of Munich, TUM School of Natural Sciences,
Physics Department, James-Franck-Str.~1, 85748 Garching, Germany}

\author[0000-0002-7040-9472]{Robert Glas}
\affiliation{Max Planck Institute for Astrophysics, Karl-Schwarzschild-Str.~1, 85748, Garching, Germany}

\author[0000-0002-0831-3330]{H.-Thomas Janka}
\affiliation{Max Planck Institute for Astrophysics, Karl-Schwarzschild-Str.~1, 85748, Garching, Germany}

\author[0000-0003-2912-9978]{Jakob Ehring}
\affiliation{Institute of Physics, Academia Sinica, No.~128, Sec.~2, Academia Rd., 115201 Taipei City, Taiwan}
\affiliation{Max-Planck-Institut f\"ur Physik (Werner-Heisenberg-Institut), Boltzmannstr.~8, 85748 Garching, Germany}
\affiliation{Max Planck Institute for Astrophysics, Karl-Schwarzschild-Str.~1, 85748, Garching, Germany}

\author[0000-0001-5481-7727]{Ernazar Abdikamalov}
\affiliation{Department of Physics and Energetic Cosmos Laboratory, Nazarbayev University, Astana 010000, Kazakhstan}

\author[0000-0002-3126-9913]{Oliver Just}
\affiliation{GSI Helmholtzzentrum für Schwerionenforschung, Planckstr.~1, 64291 Darmstadt, Germany}
\affiliation{Astrophysical Big Bang Laboratory, RIKEN Cluster for Pioneering Research, 2-1 Hirosawa, Wako, Saitama 351-0198, Japan}



\begin{abstract}
Accretion-induced collapse (AIC) or merger-induced collapse (MIC) of white dwarfs (WDs) in binary systems is an interesting path to neutron star (NS) and magnetar formation, alternative to stellar core collapse and NS mergers. Such events could add a population of compact remnants in globular clusters, they are expected to produce yet unidentified electromagnetic transients including gamma-ray and radio bursts, and to act as sources of trans-iron elements, neutrinos, and gravitational waves. Here we present the first long-term ($\gtrsim$\,5\,s post bounce) hydrodynamical simulations in axi-symmetry (2D), using energy- and velocity-dependent three-flavor neutrino transport based on a two-moment scheme. Our set of six models includes initial WD configurations for different masses, central densities, rotation rates, and angular momentum profiles. Our simulations demonstrate that rotation plays a crucial role for the proto-neutron star (PNS) evolution and ejecta properties. We find early neutron-rich ejecta and an increasingly proton-rich neutrino-driven wind at later times in a non-rotating model, in agreement with electron-capture supernova models. In contrast to that and different from previous results, our rotating models eject proton-rich material initially and increasingly more neutron-rich matter as time advances, because an extended accretion torus forms around the PNS and feeds neutrino-driven bipolar outflows for many seconds. AIC and MIC events are thus potential sites of r-process element production, which may imply constraints on their occurrence rates. Finally, our simulations neglect the effects of triaxial deformation and magnetic fields, yet they provide valuable reference cases for comparison with future long-term magneto-hydrodynamic and three-dimensional AIC studies.
\end{abstract}



\section{Introduction} \label{sec:intro}

Accretion-induced collapse (AIC) or merger-induced collapse (MIC) of white dwarfs (WDs) near the Chandrasekhar mass limit or at super-Chandrasekhar masses for rotationally stabilized WDs were proposed long ago and have repeatedly been considered as alternative paths to neutron star (NS) formation (e.g., \citealt{Nomoto+1979,Canal+1980a,Canal+1980b,Miyaji+1980,Isern+1983,Miyaji+1987,Nomoto+1991,Canal+1990,Isern+1991,Freire+2014,Schwab+2015,Schwab+2016}; see also \citealt{Postnov+2014} for a review). Such events, which we often collectively call AICs here, could happen in cases where ONe WDs receive mass from a companion star in a close binary system either by Roche-lobe overflow or in a common-envelope scenario, or when they are left as remnants of the merger of two WDs. These evolution paths are likely to lead to rapidly spinning, centrifugally supported, degenerate objects \citep{Piersanti+2003a,Piersanti+2003b,Uenishi+2003,Saio+2004,Yoon+2004,Yoon+2005,Kashyap+2017,Kashyap+2018,Ablimit+2022}. The collapse is possible if the accreting or merging WDs get around disruption by thermonuclear burning, i.e., they need to avoid producing a Type~Ia supernova (SN) or stripping significant parts of their mass in any other kind of thermonuclear runaway process, for example a thermonuclear electron-capture SN \citep{Jones+2016,Jones+2019}. 

If ONe deflagration starts only at high densities or is inefficient, the gravitational collapse is triggered by the onset of electron captures mainly on $^{20}$Ne when the density approaches about $10^{10}$\,g\,cm$^{-3}$ \citep{Kirsebom+2019,Zha+2019,Suzuki+2019}. Such events are expected to produce multi-messenger signals including neutrinos and gravitational waves \citep{Fryer+2002,Abdikamalov+2010,LongoMicchi+2023}, and since they eject fractions of a solar mass in wind-like outflows driven by neutrino heating \citep{Woosley+1992} that might be viable for neutron-rich nucleosynthesis \citep{Fryer+1999,Dessart+2006,Yip+2024}, they could lead to observationally faint electromagnetic transients \citep{Nomoto1986}, potentially also to long-duration gamma-ray bursts (GRBs) accompanied by kilonovae \citep{ChiKitCheong+2024}. Moreover, they were discussed as explanations for the properties of pulsars in X-ray binaries and as origin of the large populations of millisecond pulsars in globular clusters \citep[e.g.,][]{vandenHeuvel1984,Taam+1986,Bailyn+1990,Bhattacharya+1991,Ivanova+2008,Hurley+2010,Boyles+2011,Lynch+2012,Freire+2014,Kremer+2023}, as potential sources of fast radio bursts \citep{Margalit+2019,Kremer+2023}, and as birth events of low-mass NSs \citep{Leung+2019}.

The occurrence rate of AIC and MIC events is highly uncertain but could reach a fair fraction of the Type Ia SN rate. The most optimistic estimates based on binary population synthesis predict rates in the Milky Way of $\sim$$(0.3-0.9)\times 10^{-3}$\,yr$^{-1}$ for AIC via ONe WD+He star and CO WD+He star single-degenerate channels \citep{Wang2018} and of $(1.4-8.9)\times 10^{-3}$\,yr$^{-1}$ for MIC when all ONe/CO WD+ONe/CO WD double-degenerate mergers are taken into account \citep{Liu+2020,WangLiu2020}. Although AIC events have not been unambiguously identified yet, there are several candidate events whose observational properties cannot be explained by conventional scenarios of stellar death \citep[e.g.][]{Moriya2019,McBrien+2019,Gillanders+2020}. Moreover, the number of discovered possible candidates of AIC progenitors is growing, including, e.g., super-Chandrasekhar objects as products of ONe+CO WD mergers \citep{Oskinova+2020}, rapidly rotating, highly magnetized WDs close to the Chandrasekhar mass limit \citep{Caiazzo+2021}, or binary systems with the properties to merge to super-Chandrasekhar WDs \citep{Luo+2024,Zhao+2024}.     

To clearly link peculiar astronomical transients to NS formation from collapsing WDs, using their light-curve, spectral, and chemical characteristics, better theoretical models of this phenomenon are needed. Predictions of such observables require detailed neutrino-hydrodynamic and magneto-hydrodynamic (MHD) simulations that permit reliable calculations of ejecta mass, geometry, and composition, and thus of the nucleosynthesis and electromagnetic emission. Instead, \citet{Darbha+2010} computed light curves and spectra based on a parametric description of AIC outflows, which were assumed to be spherically symmetric (1D) and rich of $^{56}$Ni and iron-group elements, motivated by simplified (height-integrated, time-dependent) models for viscously driven evolution of accretion disks around rotating NSs \citep{Metzger+2009}. 

AICs have been investigated by neutrino-hydrodynamical simulations repeatedly, but the corresponding models are still limited with respect to their sophistication and evolution periods. \citet{Woosley+1992} performed 1D general relativistic (GR) simulations to study the ejection of a baryonic wind that is driven by neutrino heating near the proto-NS (PNS) surface; they discussed possible implications for GRBs, nucleosynthesis, rates, and detectable signals. More modern versions of such 1D simulations, considering sets of constructed hydrostatic initial WD configurations instead of an evolutionary stellar core model, were recently presented by \citet{Mori+2023} and \citet{Yip+2024}. \citet{Fryer+1999} conducted a large sample of calculations in 1D and 2D (i.e., with axi-symmetry including one rotating model and with variations of other ingredients), using smoothed-particle hydrodynamics and gray (i.e., energy integrated) neutrino transport with neutrino interaction rates that are no longer compatible with state-of-the-art treatments. Similarly, the 2D Newtonian hydrodynamic and MHD simulations reported by \citet{Dessart+2006} and \citet{Dessart+2007}, respectively, used multi-group flux-limited diffusion with various other approximations in the transport (e.g., no Doppler-velocity terms) and in the neutrino rates. \citet{Abdikamalov+2010} compared a large grid of 2D GR simulations for AICs of WDs with different rotation, however without neutrino transport and a strong focus on the gravitational-wave signals around core bounce and shortly afterwards. In a similar way, also sacrificing neutrino transport, \citet{Zha+2019a}, \citet{Leung+2019}, and \citet{Chan+2023} studied the consequences of dark matter admixture in WDs on the AIC dynamics and GW emission, and \citet{LongoMicchi+2023} and \citet{ChiKitCheong+2024} performed 3D GR hydrodynamic and 2D GR-MHD simulations, respectively, with gray M1 neutrino transport (and grid resolutions on their finest adaptive mesh-refinement levels of 0.369\,km and 0.488\,km, respectively, whereas our models use 0.1\,km and test-wise even 0.03\,km to resolve steep radial PNS density gradients) to estimate multi-messenger AIC signals (gravitational waves, neutrinos, electromagnetic transients such as GRBs and kilonovae) based on roughly 150\,ms of post-bounce evolution in 3D and up to $\sim$2\,s in 2D. Thus most of these simulations covered evolution periods of just fractions of a second after bounce and/or used approximations in the neutrino transport and interactions with unclear implications for the predictions of nucleosynthesis conditions in the ejecta.

Dynamically, the outcome of such simulations without rotation is similar to the electron-capture induced collapse of degenerate ONe cores in progenitors of so-called electron-capture supernovae (ECSNe), yielding low-energy explosions (i.e., explosion energies $E_\mathrm{exp}$ around or less than $10^{50}$\,erg) and a small amount of neutrino-heated ejecta that carry iron-group and some trans-iron nuclei \citep{Kitaura+2006,Wanajo+2009,Wanajo+2011,Wanajo+2018}. Although ECSN explosions can be obtained in 1D models, multi-dimensional simulations are indispensable to obtain quantitatively meaningful results for the observables \citep{Janka+2008,Wanajo+2011,Melson+2015,Radice+2017,Stockinger+2020,Wang+2024}.   

In this paper we present the first long-term ($\gtrsim$\,5\,s post bounce) simulations of AICs in 2D for initial WD configurations with different masses, rotation rates, rotation profiles, and pre-collapse central densities. Here we focus on reporting the nucleosynthetic conditions in the ejecta, connected to the neutrino emission of the new-born PNSs. More details on the initial models and on the hydrodynamic evolution will be provided elsewhere. Our results of a non-rotating model are in line with previous (though shorter) multi-D simulations of the collapse of degenerate stellar cores in ECSNe, where the earliest and fastest ejecta are neutron-rich and the later ejecta become increasingly proton-rich \citep{Wanajo+2011,Wanajo+2018,Stockinger+2020}. However, rotation reverses this trend and leads to relatively slower, proton-dominated, polar outflows initially, followed by increasingly neutron-rich ejecta in growing latitudinal regions as time advances. This fundamental change emphasizes the crucial role of rotation for the nucleosynthetic output of AICs and for corresponding conclusions that can be drawn on the observable brightness \citep[e.g.,][]{Metzger+2009,LongoMicchi+2023} and on constraints of the still highly uncertain rates of such events \citep[e.g.,][]{Woosley+1992,Fryer+1999,Dessart+2007}.

Section~\ref{sec:methods} provides a brief description of the numerical code used for our neutrino-hydrodynamic simulations. It also describes the basic principles behind the construction of our initial WD models and the crucial properties of the set of six different initial WD configurations selected for our modeling project. Moreover, it contains a short elucidation of the aims as well as limitations of the current study. In Section~\ref{sec:results} we discuss our results with a focus on the nucleosynthesis-relevant ejecta properties and their explanation on grounds of the ejecta dynamics and neutrino interactions. In Section~\ref{sec:summary} we summarize our results and discuss them in relation to previous work.

\begin{deluxetable*}{lcccrcrcccccc}\label{tab:pars}
\setlength{\tabcolsep}{3pt}
    \tablecaption{Properties of the AIC progenitors. $M_\mathrm{WD}$ is the baryonic mass of the initial WD, $\Omega_\mathrm{c}$ is its central angular velocity, $\Omega_\mathrm{max}$ is its maximum value of the angular velocity at the equatorial radius $R_\mathrm{max}$, $\Omega_\mathrm{s}$ is its angular velocity (95\% of the Keplerian mass-shedding limit) at the equatorial surface radius$R_\mathrm{eq}$, $R_\mathrm{eq}R_\mathrm{pole}^{-1}$ is its ratio of the surface radii at equator and pole, $\rho_\mathrm{c}$ is its central density, $T_\mathrm{c}$ is its central temperature, $J_\mathrm{WD}$ is its total angular momentum, $E_\mathrm{rot,WD}$ is its total rotational energy, and $\beta_\mathrm{i}$ is its ratio of rotational to gravitational energy, all given for the initial WD prior to collapse.
}
    \setlength{\tabcolsep}{4pt}
    \tablehead{
        Model &  $M_\mathrm{WD}$  & $\Omega_\mathrm{c}$ & $\Omega_\mathrm{max}$ & $R_\mathrm{max}$ & $\Omega_\mathrm{s}$ & $R_\mathrm{eq}$ & ${\displaystyle \phantom{\int\hspace{-10pt}}\frac{R_\mathrm{eq}}{R_\mathrm{pole}} }$ & $\rho_\mathrm{c}$ & $T_\mathrm{c}$  & $J_\mathrm{WD}$  & $E_\mathrm{rot,WD}$   & $\beta_\mathrm{i}$\\
        & [M$_\odot$] & ${\displaystyle\mathrm{\left[\frac{rad}{s}\right]}}$ & ${\displaystyle\mathrm{\left[\frac{rad}{s}\right]}}$ & $\mathrm{[km]}$ & {$\displaystyle\mathrm{\left[\frac{rad}{s}\right]}$} & $\mathrm{[km]}$  &     $\phantom{\int\limits_{.}}$                      & $\left[10^9\,{\displaystyle\mathrm{\frac{g}{cm^3}}}\right]$ & $\left[10^9\,\mathrm{K}\right]$  & $\left[10^{50}\,\mathrm{erg\,s}\right]$  & $\left[10^{50}\,\mathrm{erg}\right]$ &  }
        \startdata
        M1.42-J0       &1.422 &0 &0 & --- & 0 &816 &1.00 & 50.0 & 10.0 &0.000 &0.000 &0.000      \\
        M1.42-J0.23-Dl &1.422 &0.002 &3.80 & 2250 & 3.80 &2250 &1.34 &4.0  & 4.13 &0.231 &0.299 &0.008 \\
        M1.61-J0.47 &1.609 &12.01 &19.65 & 411 & 7.53 &1498 &2.04 &50.0 &10.0 &0.471 &3.941 &0.038   \\
        M1.61-J0.78-Dl &1.611 &5.55 &6.13 & 801 & 2.79 &2897 &1.93 &4.0 &4.13 &0.782 &2.094 &0.046  \\
        M1.91-J1.09 &1.919 &18.01 &25.67 & 446 & 4.14 &2377 &3.55 &50.0 &10.0 &1.085 &10.471 &0.082  \\
        M1.91-J1.63-Dl &1.906 &5.34 &8.20 & 864 & 1.88 &3982 &2.87 &4.0 &4.13&1.630 &5.033 &0.091   \\
       \enddata
\end{deluxetable*}

\section{Methods} 
\label{sec:methods}

\subsection{Numerical Code}
\label{sec:code}

To conduct our 2D simulations, we employed the neutrino-hydro\-dynamics code \textsc{Alcar} \citep{Obergaulinger2008,Just+2015,Just+2018}. The code couples Newtonian hydrodynamics with energy-dependent transport for neutrinos of all three flavors ($e, \mu, \tau$), based on a two-moment scheme with an analytical closure relation. Gravitational effects are taken into account by solving the Poisson equation for a self-gravitating fluid in 2D and by including GR corrections in the monopole term of the gravitational potential according to Case~A of \citet{Marek+2006} as well as general relativistic redshift and time dilation corrections in the transport equations, following \citet{Rampp+2002}.
 
The hydrodynamics module integrates the Euler equations for the conservation of mass, momentum, energy, and electron-lepton number, including the corresponding source terms (of momentum, energy, lepton number) for neutrinos. The corresponding solver uses a Godunov-type, finite-volume scheme in spherical polar coordinates with higher-order reconstruction methods and an approximate Riemann solver (HLLC, except in regions of grid-aligned shocks, where the more diffusive HLLE solver is applied) to obtain cell-interface fluxes. The time integration is performed by an explicit, second-order Runge-Kutta scheme with the time step being constrained by the Courant-Friedrichs-Lewy (CFL) condition \citep{Courant+1928}. 

The fully 2D neutrino transport is solved for three neutrino species, i.e., electron neutrinos $\nu_e$, electron antineutrinos $\bar{\nu}_e$, and $\nu_x$, which collectively represents the heavy-lepton neutrinos ($\nu_\mu, \bar{\nu}_\mu, \nu_\tau, \bar{\nu}_\tau $, whose opacities are basically identical apart from minor details). The code evolves the energy density and energy-flux density in the comoving frame of the stellar plasma by conservative, mixed explicit and implicit time integration of the zeroth- and first-order angular moment equations of the Boltzmann equation, which are closed by Minerbo's algebraic relations for the higher-order radiation moments, e.g., in the Eddington tensor \citep[``M1'' scheme;][]{Minerbo1978}. The hyperbolic transport equations are numerically integrated by the same methods as applied for the hydrodynamics module, using a Godunov-type, finite-volume scheme. Velocity-dependent terms up to order $v/c$ and Doppler and gravitational energy shifting are included. The neutrino-matter interactions through weak charged-current and neutral-current reactions with nucleons, nuclei, electrons, and positrons, as well as neutrino-pair processes (electron-positron pair annihilation and nucleon-nucleon bremsstrahlung) are taken into account, partially via implicitly treated source terms in the moment equations. For details on the numerical aspects of the \textsc{Alcar} code, see \citet{Just+2015}, and for more recent updates of the microphysics and neutrino reactions, see \citet{Just+2018}. The exact setup for the neutrino interactions used in our AIC simulations reported here is described in \citet{Glas+2019}. 

\textsc{Alcar} contains all the neutrino transport physics for $\nu_e$, $\bar\nu_e$, and $\nu_x$ to treat electron and positron captures and their inverse neutrino absorptions, neutrino-pair processes, and neutrino-scattering reactions off electrons, positions, nucleons, and nuclei during the collapse of degenerate stellar cores and during the subsequent post-bounce and PNS evolution. This makes it unnecessary to employ the simple parametrization proposed by \citet{Liebendoerfer2005}, which was often used for approximating the deleptonization prior to core bounce \citep[e.g., in][]{Abdikamalov+2010,LongoMicchi+2023,ChiKitCheong+2024}. However, this parametrization in terms of a prescribed relation $Y_e(\rho)$ for the electron fraction $Y_e$ as a function of the density $\rho$ was deduced from spherically symmetric core-collapse simulations. Therefore its validity in the collapse of rapidly rotating models is uncertain. We ignore neutrino flavor oscillations in our treatment, thus adopting the standard assumption in stellar core-collapse modeling. 

The code is supplemented by an equation of state (EoS) for the thermodynamics and composition of the stellar plasma, which assumes the composition to be described by nuclear statistical equilibrium (NSE) at densities $\rho \ge 2000$\,g\,cm$^{-3}$, where we use the tabulated SFHo EoS of \citet{Steiner+2013} (binding energy of symmetric matter at nuclear saturation density: 16.19\,MeV; incompressibility: 245\,MeV; symmetry energy: 31.57\,MeV; slope of the symmetry energy: 47.10\,MeV). It is available for electron fractions $Y_e \le 0.6$, and temperatures $T \ge 0.1$\,MeV, but it was extended towards higher $Y_e$ up to 0.9 and lower temperatures down to $10^{-3}$\,MeV in the density regime below $10^{11}$g\,cm$^{-3}$ (where nucleon interactions play a negligible role). At $\rho < 2000$\,g\,cm$^{-3}$, the stellar plasma is treated by a combination of ideal gases of photons, arbitrarily degenerate and arbitrarily relativistic electrons and positrons, as well as contributions from Boltzmann gases of non-relativistic neutrons, protons, alpha particles, and a representative, fixed heavy nucleus in NSE above temperatures of 0.43\,MeV. Below $T = 0.43$\,MeV the nuclear composition is assumed to be frozen in this density regime and advected with the stellar fluid. This description is applicable for any $Y_e$ and down to very low temperatures and densities \citep{Janka+1995,Janka+1996}. It takes into account the effects of nuclear dissociation and recombination and is a sufficiently good approximation for the thermodynamics of the stellar plasma, which is dominated by the contributions of the charged leptons and photons. The stored nuclear (NSE) composition in the high-density domain, where neutrino interactions with the stellar medium have to be computed, includes neutrons, protons, deuterons, tritons, helium-3, alpha particles, and an average heavy nucleus, which depends on $\rho$, $T$, and $Y_e$.

For the neutrino energy, we use 15 logarithmically spaced bins in the energy range from 0 to 400\,MeV for all neutrino species. Our spatial grid consists of 1440 radial zones and is constructed with equidistant cell widths of 0.1\,km from the grid center ($r = 0$) up to 30\,km and with logarithmically increasing cell widths up to 600,000\,km. The angular grid uses 96 equidistant angular zones, resulting in an angular resolution of 1.875$^{\circ}$ for $0\le \theta \le 180^\circ$. Since after several seconds of post-bounce evolution the density gradient in a layer containing the PNS's surface can become extremely steep, a remapping routine was implemented into the \textsc{Alcar} code, which interpolates the fluid and neutrino data to a new grid with higher radial resolution while keeping the angular mesh unchanged. For some of our AIC models we tested the convergence of the results with respect to the radial resolution by repeating parts of these simulations with a new radial grid with 1600 zones. The 160 additional grid cells are distributed such that we obtain a resolution of 0.03\,km between 9\,km and 15\,km.

In order to ease the CFL time step constraint, the innermost core is treated in spherical symmetry (1D). This central 1D region has a radius of 1\,km when simulating the collapse of non-rotating WD models, 600\,m for slowly rotating WD models, and 400\,m for rapidly rotating WD models. Conservation of rotational angular momentum is ensured by allowing the 1D core to rotate with non-zero $\phi$-velocities, corresponding to the angular momentum of matter advected into the 1D core. The numerically motivated use of this small 1D core has no relevant impact on the simulations.

\subsection{Progenitor Models}
\label{sec:progenitors}

Since self-consistent 2D WD models from stellar evolution or WD merger calculations, in particular with different amounts of rotation, are not available at the pre-collapse stage, we constructed our initial WD progenitors as Newtonian configurations in rotational equilibrium with the self-consistent field (SCF) method \citep{Ostriker+1968a,Ostriker+1968b,Komatsu+1989a,Komatsu+1989b}, following \citet{Yoon+2005} and \citet{Abdikamalov+2010}. For a barotropic, rotating star \citep[i.e., pressure $P = P(\rho)$ and $\partial\Omega/\partial z = 0$ for the angular velocity, where $z = r\cos\theta$ is the distance to the equatorial plane;][]{Tassoul1978} with chosen central density $\rho_\mathrm{c}$ and central angular velocity $\Omega_\mathrm{c}$, the SFC method iterates the 2D density distribution $\rho(r,\theta)$ to convergence at a defined accuracy. For doing so, it uses a constructed rotation law based on realistic physical assumptions, i.e., a physically constrained functional behavior of with $\Omega(s)$ ($s = r\sin\theta$ being the distance from the rotation axis) as detailed in \citet{Yoon+2004}, \citet{Yoon+2005}, and \citet{Abdikamalov+2010}. The EoS of the initial WDs is assumed to be given by a completely degenerate electron gas \citep{Ostriker+1968b}, for which $P = P(\rho)$ applies, and the electron fraction is set to be constant at a value of $Y_e = 0.5$. 

The mass of our WD progenitors was determined by the chosen initial values of $\rho_\mathrm{c,i}$, $\Omega_\mathrm{c,i}$, and the angular velocity profile, which reached an absolute maximum $\Omega_\mathrm{max}$ at an equatorial radius where the density was chosen to be a fraction of 5\% of the initial central density for our models with a high value of $\rho_\mathrm{c,i} = 5\times 10^{10}$\,g\,cm$^{-3}$ and of 10\% for our models with a low value of $\rho_\mathrm{c,i} = 4\times 10^9$\,g\,cm$^{-3}$. All of our initial WD models with non-zero angular momentum rotate with 95\% of the Keplerian angular velocity at the surface on the equator, which is a consequence of their imagined accretion or merger evolution in binary systems as argued by \citet{Yoon+2004} and \citet{Yoon+2005}, who justified the considered rotation rates and angular momentum profiles for rapidly accreting WDs \citep[accretion rates $>$10$^{-7}$\,M$_\odot$\,yr$^{-1}$; for a summary, see also][]{Abdikamalov+2010}. The models were assumed to be converged when the relative changes of the iterated quantities dropped below $10^{-5}$ on the entire 2D grid. They were constructed with high spatial ($r$-$\theta$) resolution such that in all models at least 97\% of the WD mass fulfilled rotational equilibrium (i.e., the balance of gravitational, pressure, and centrifugal accelerations) with a relative accuracy better than 5\% of the local gravitational acceleration \citep[see][]{Ehring2019}. Fulfilling these criteria, our initial WD models are more accurate than those used in previous applications of similarly constructed pre-collapse configurations.

When starting our AIC calculations with a non-zero temperature EoS, we imposed an initial temperature-density relation onto the WD models according to
\begin{equation}
T_\mathrm{i}(r,\theta) = T_0\left[\frac{\rho_0}{\rho_\mathrm{i}(r,\theta)}\right]^{-0.35}
\label{eq:tempini}
\end{equation}
with $T_0 = 10^{10}$\,K and $\rho_0 = 5\times 10^{10}$\,g\,cm$^{-3}$ \citep[for similar treatments, see][but note an error in the sign of the exponent in the corresponding equations there]{Dessart+2006,Abdikamalov+2010}. This prescription means that the central temperatures $T_\mathrm{c,i}$ in the initial WD models differ, depending on their different central densities $\rho_\mathrm{c,i}$ (see Table~\ref{tab:pars}). Because of the high initial value of $Y_e$, electron captures set in immediately and triggered the collapse in all cases.

Because our AIC simulations were performed with a grid-based code, we had to fill the computational volume exterior to the WDs with a low-density circumstellar medium (CSM). For that we imposed power-law density and temperature profiles as functions of radius $r$ and polar angle $\theta$: $\rho(r,\theta) = \rho_\mathrm{WD,min}(R_\mathrm{WD}(\theta)/r)^{3+x}$ and $T(r,\theta) = T_\mathrm{WD,min}R_\mathrm{WD}(\theta)/r$ with $x = 0.1$ and $\rho_\mathrm{WD,min}$ and $T_\mathrm{WD,min}$ being the minimum density and temperature at the angle-dependent radius $R_\mathrm{WD}(\theta)$ of the WD surface, respectively. This surrounding medium was assumed not to rotate, and the radial density decline ensured a finite integrated mass even for large radii of the outer grid boundary. Moreover, the assumed radial density and temperature profiles allowed the CSM to be close to hydrostatic equilibrium independent of whether its pressure is dominated by Boltzmann gases or radiation.\footnote{This, however, is not perfectly fulfilled because in general the WD is not a spherical gravitating body and the true EoS of the medium is more complex.} We set the electron fraction in the CSM to a constant value of $Y_{e,\mathrm{CSM}} = 0.6$, determined by this upper limit of the initial EoS table. (The EoS table was extended in $Y_e$ to higher values only later, after the AIC simulations had developed ejecta with $Y_e$ exceeding the initial upper limit of the table.) In practice, the chosen CSM density turned out to be too large, competing with the ejecta mass in some of our models, and the $Y_e$ value is too low, because explosive outflows can develop similarly high and even larger $Y_e$, making a discrimination of WD ejecta and CSM difficult. We made sure that these caveats do not affect the results reported here, but in future simulations that follow the ejecta to larger radii, the maximum CSM density at the WD surface will be lowered and $Y_{e,\mathrm{CSM}}$ will be set to a higher value.  

For a first survey of the relevant parameter space of initial conditions, we studied a set of six well-resolved WD models, which bracket the spectrum of possibilities, covering different masses, central densities, and spins and angular momentum distributions, including differentially and nearly rigidly rotating models (see Table~\ref{tab:pars}). Model M1.42-J0 is a non-rotating case with high initial central density and M1.42-J0.23-Dl a slowly rotating model with low initial central density (indicated by the suffix ``Dl'' of its name), both with baryonic masses of 1.422\,M$_\odot$ near the Chandrasekhar limit. The initial baryonic mass of the WD (in M$_\odot$) is denoted by the number following the letter ``M'' of the model name, and the value of the WD's total angular momentum (in $10^{50}$\,erg\,s) is given by the number following the letter ``J''. The other investigated cases are rotationally stabilized super-Chandrasekhar models, which could be imagined as WD-merger products. M1.61-J0.78-Dl and M1.91-J1.63-Dl are nearly uniformly rotating models with low initial central densities and masses of about 1.61\,M$_\odot$ and 1.91\,M$_\odot$, respectively, and M1.61-J0.47 and M1.91-J1.09 are high-density counterparts with masses very close to these two values but more extreme differential rotation.

\subsection{AIC Modeling Goals}

Our main goal is studying the WD collapse and ejecta dynamics in order to determine the nucleosynthesis conditions in AIC outflows. Using our elaborate transport treatment and extending our simulations to longer evolution periods, we thus surpass the limitations of previous works. However, in the present AIC simulations we neither consider any physical viscosity nor magnetic fields, although WDs including those that accrete from binary companions can possess magnetic fields up to about $10^9$\,G \citep{Ferrario+2015}, and especially remnants of double-WD mergers are expected to be highly magnetized due to their rapid rotation and dynamo processes during the merger \citep{Caiazzo+2021}. Moreover, seed magnetic fields can be further amplified by compression, turbulent flows, dynamo activity, winding due to differential rotation, and the magnetorotational instability during the WD collapse and the subsequent PNS evolution. 

Therefore, since we ignore any effects of magnetic fields in our models, we consider them as benchmark cases and reference points for future explorations of a wider parameter space of pre-collapse WD conditions, which are currently not tightly constrained by observations and not well probed by self-consistent evolution calculations of accreting WDs and remnants of WD mergers. Uncertainties in the initial magnetic field strength and configuration enlarge the space of possibilities for the WD progenitor models even more. In any case, investigations of the non-magnetic conditions are indispensable to comprehend the additional physical effects connected to MHD. Our purely hydrodynamic models are therefore a well defined and solid basis for adding more complexity in the future. 

Another constraint applies to our restriction to 2D calculations, which permits us to extend our AIC simulations with high radial resolution over evolution periods of several seconds by employing sizable, though still acceptable, computational resources. Such long-term simulations for a larger set of cases are currently not feasible in three dimensions (3D), though they would be indispensable for reliable MHD results because of generic 3D phenomena connected to magnetic fields. In this context, triaxial deformation that is triggered by purely hydrodynamical non-axisymmetric (spiral-mode and bar-mode) instabilities are another effect that is missed in 2D simulations. While the ratios of rotational energy to gravitational energy, $\beta \equiv {\cal T}/|W|$, of all of our initial WD models ($\beta_\mathrm{i}$, see Table~\ref{tab:pars}) are well below the critical limits of ${\cal T}/|W|\gtrsim 0.14$ for secular instability and ${\cal T}/|W|\gtrsim 0.26$ for dynamical instability \citep{Shapiro+1983}, the situation changes after bounce, at which time the lower critical value is already approached in the unshocked cores of our rapidly rotating, low-density WD models (M1.61-J0.78-Dl, M1.91-J1.63-Dl). Because of the highly differentially rotating states, a low-${\cal T}/|W|$ instability with one-armed spiral or bar-like asymmetry modes may be expected to occur on dynamical timescales even for significantly lower values of ${\cal T}/|W|$ \citep[e.g.,][]{Tohline+1990,Shibata+2002,Shibata+2003,Centrella+2001,Saijo+2003,Ott+2005,Saijo+2006,LongoMicchi+2023}.  

Despite these shortcomings with respect to dimensionality and magneto-viscous effects, our modeling project constitutes a next major step beyond the current status by exploring a wider range of initial WD configurations than considered previously and by following their AIC evolution for much longer periods (up to nearly 8\,s instead of fractions of a second after bounce) and with an unprecedentedly detailed treatment of the crucial neutrino effects.

\begin{deluxetable*}{lcccccccccccccccc}\label{tab:sims}
\setlength{\tabcolsep}{3pt}
    \tablecaption{Properties of the AIC ejecta and remnants. $M_\mathrm{WD}$ is the baryonic mass of the initial WD,  $t_\mathrm{b}$ is the bounce time, $M_\mathrm{ic,b}$ is the mass of the inner core at bounce, $\beta_\mathrm{ic,b}$ is the ratio of rotational to gravitational energy of the inner core at bounce, $t_\mathrm{f}$ is the post-bounce time when the AIC simulation was stopped, and $M_\mathrm{NS}$ is the corresponding baryonic mass of the NS (defined as the mass with density $\rho > 10^{10}$\,g\,cm$^{-3}$), $M_\mathrm{tor}$ is the torus mass, $\beta_\mathrm{rem,f}$ is the ratio of rotational to gravitational energy of $M_\mathrm{NS}+M_\mathrm{tor}$ (i.e., of the gravitationally bound remnant consisting of NS plus torus), $E_\mathrm{ej}$ is the energy of ejected outflows (in parentheses for the assumption that the entire ejecta assemble into a representative iron-group nucleus), $M_\mathrm{ej}$ is the ejecta mass, and $\langle Y_e\rangle_\mathrm{ej}$, $\langle s\rangle_\mathrm{ej}$, and $\langle v_r\rangle_\mathrm{ej}$ are the average electron fraction, entropy per baryon, and radial velocity, respectively, of the ejecta. The torus mass is defined as the mass with density $\rho \le 10^{10}$\,g\,cm$^{-3}$ and total energy (internal plus gravitational plus kinetic) less than or equal to zero and $Y_e \le 0.55$ or $Y_e \le 0.52$ (no relevant difference).
}
    \setlength{\tabcolsep}{4pt}
    \tablehead{
        Model &  $M_\mathrm{WD}$  & $t_\mathrm{b}$ & $M_\mathrm{ic,b}$ & $\beta_\mathrm{ic,b}$ & $t_\mathrm{f}$ & $M_\mathrm{NS}$ & $M_\mathrm{tor}$ & $\beta_\mathrm{rem,f}$ & $E_\mathrm{ej}$ & $M_\mathrm{ej}$ & $\langle Y_e\rangle_\mathrm{ej}$ & $\langle s\rangle_\mathrm{ej}$ & $\langle v_r\rangle_\mathrm{ej}$ \\
        & [M$_\odot$]  & [s] & [M$_\odot$] &   & [s]  & [M$_\odot$] & $[0.1\,\mathrm{M}_\odot]$ &   & $\left[10^{50}\,\mathrm{erg}\right]$  & $[10^{-3}\,\mathrm{M}_\odot]$ &    $\phantom{\int\limits_{.}^{1}}$ & $[k_\mathrm{B}/\mathrm{by}]$ & $\left[10^9\,{\displaystyle\mathrm{\frac{cm}{s}}}\right]$ }
        \startdata
        M1.42-J0       &1.422 & 0.0377 & 0.584 & 0.000 & 6.74 &1.409 & 0.00 & 0.000 &1.118 (1.117) &7.78 & 0.496 & 22.13 & 2.02 \\
        M1.42-J0.23-Dl &1.422 & 0.3460 & 0.541 & 0.027 & 7.83 &1.318 & 0.56 & 0.147 &2.500 (2.447) & 51.08 & 0.430 & 13.33 & 1.66 \\
        M1.61-J0.47    &1.609 & 0.0396 & 0.586 & 0.092 & 6.80 &1.468 & 1.14 & 0.263 &1.763 (1.764) & 25.12 & 0.441 & 17.97 & 2.03 \\
        M1.61-J0.78-Dl &1.611 & 0.4263 & 1.034 & 0.123 & 4.67 &1.308 & 2.85 & 0.341 &0.414 (0.438) &20.96 & 0.486 & 9.56 & 1.19 \\
        M1.91-J1.09    &1.919 & 0.0408 & 0.713 & 0.104 & 6.15 &1.471 & 4.25 & 0.310 &0.993 (1.052) & 18.66 & 0.469 & 18.29 & 1.70 \\
        M1.91-J1.63-Dl &1.906 & 0.8248 & 0.922 & 0.130 & 6.09 &1.163 & 7.45 & 0.333 &0.142 (0.221) &5.56 & 0.425 & 18.12 & 1.09 \\
       \enddata
\end{deluxetable*}

\begin{figure*}
    \centering
    \includegraphics[width=1\linewidth]{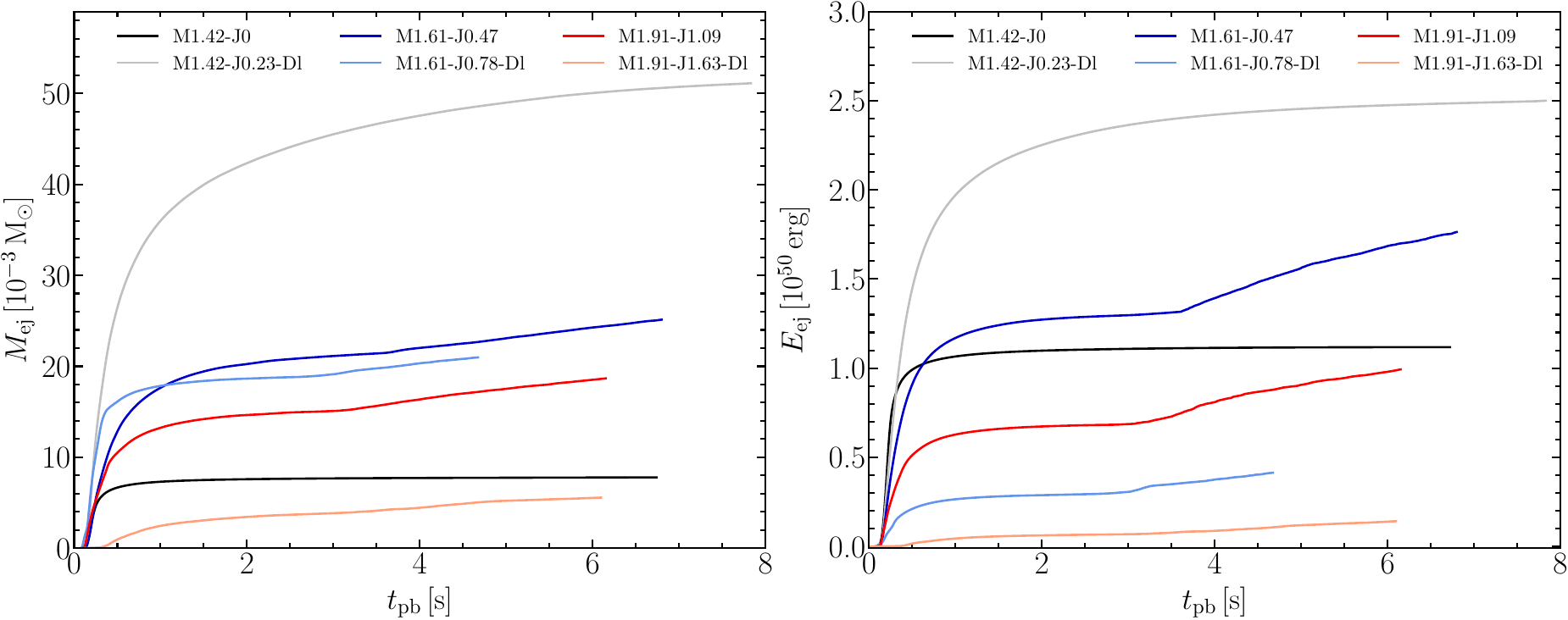}
    \caption{Cumulative ejecta masses (\textit{left}) and total (kinetic plus internal plus gravitational) ejecta energies $E_\mathrm{ej}$ (\textit{right}) as functions of time for all six AIC models. Both ejecta properties are evaluated at the outflow-evaluation sphere or spheroids, which are placed very closely outside the surfaces of the initial WDs. The steepening of the growth of the ejecta energies at $t_\mathrm{pb} > 3$\,s in the rotating models correlates with a faster rise of ejecta masses.}
    \label{fig:outflow-mass}
\end{figure*}

\begin{figure*}
    \centering
    \includegraphics[width=0.32\textwidth]{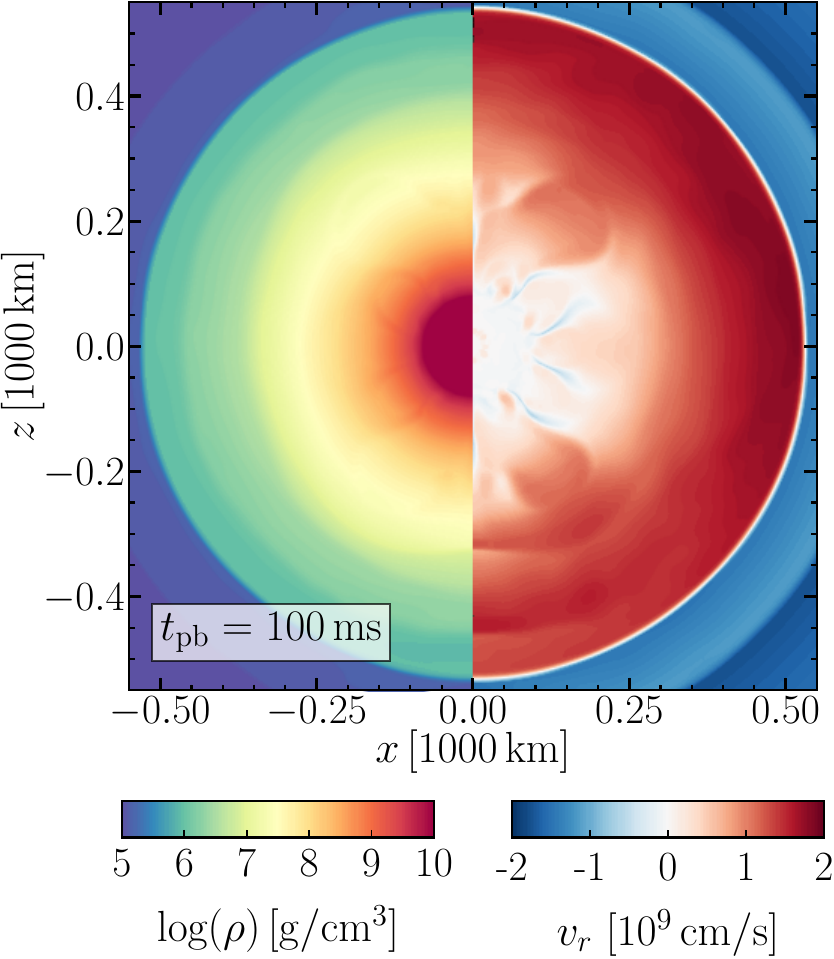}
    \includegraphics[width=0.325\textwidth]{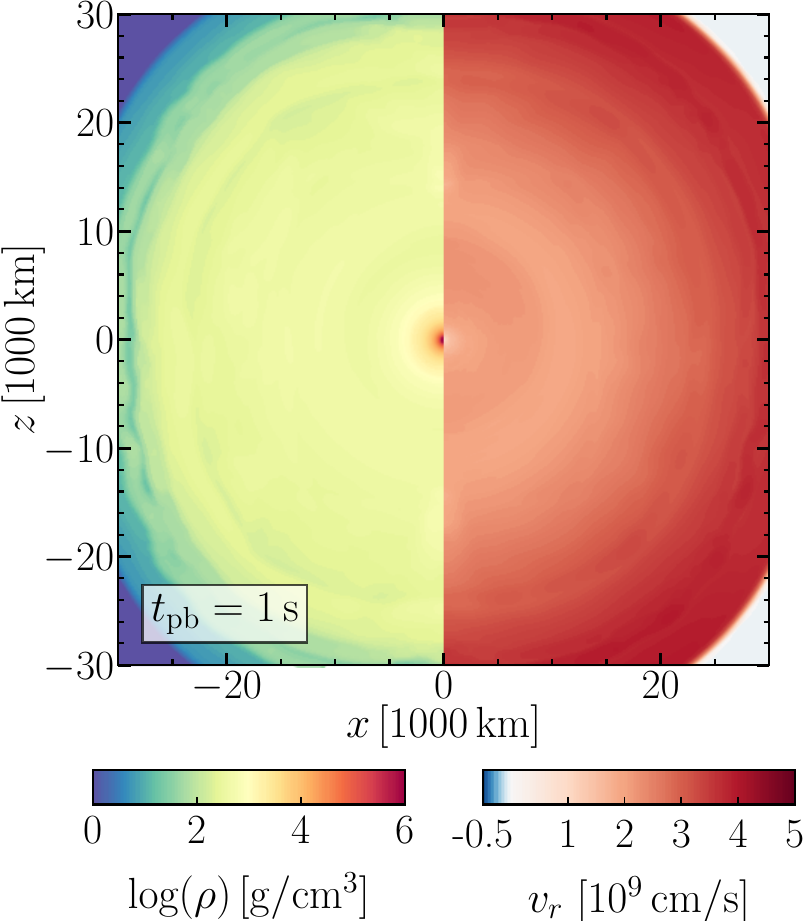}
    \includegraphics[width=0.33\textwidth]{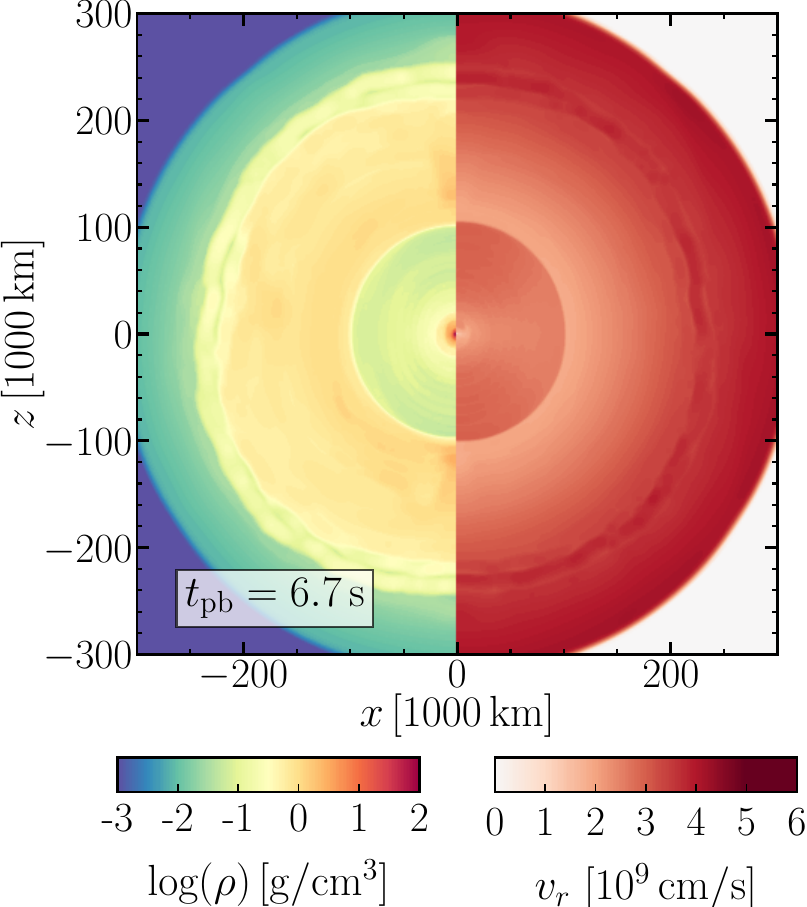}\\
    \includegraphics[width=0.325\textwidth]{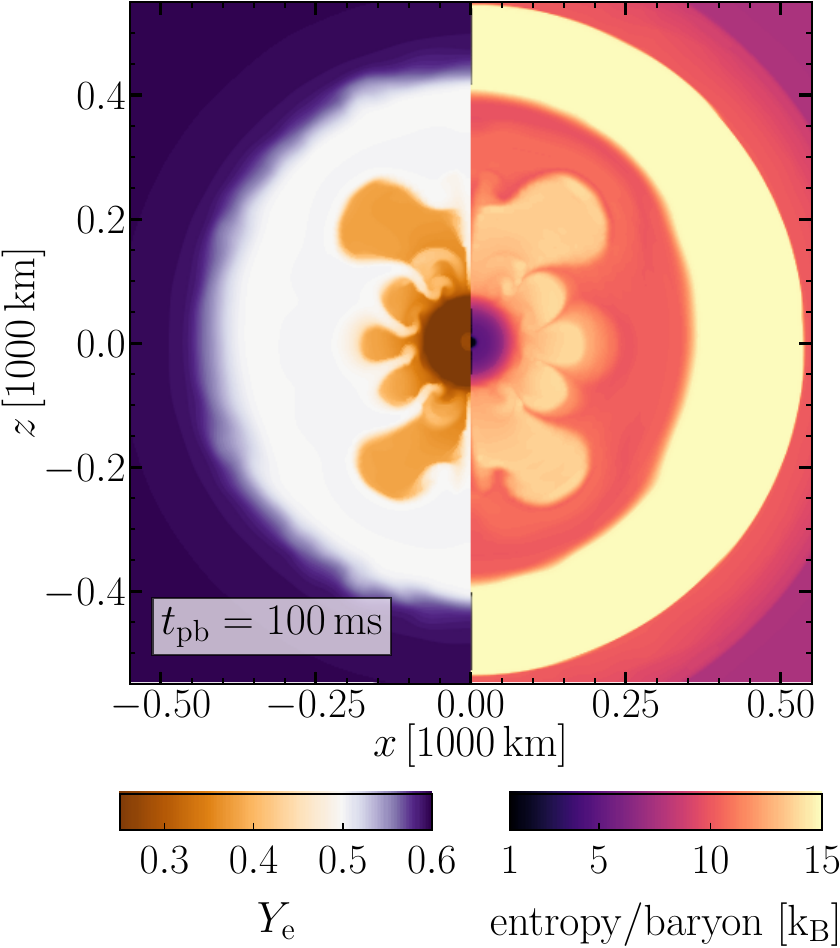}
    \includegraphics[width=0.33\textwidth]{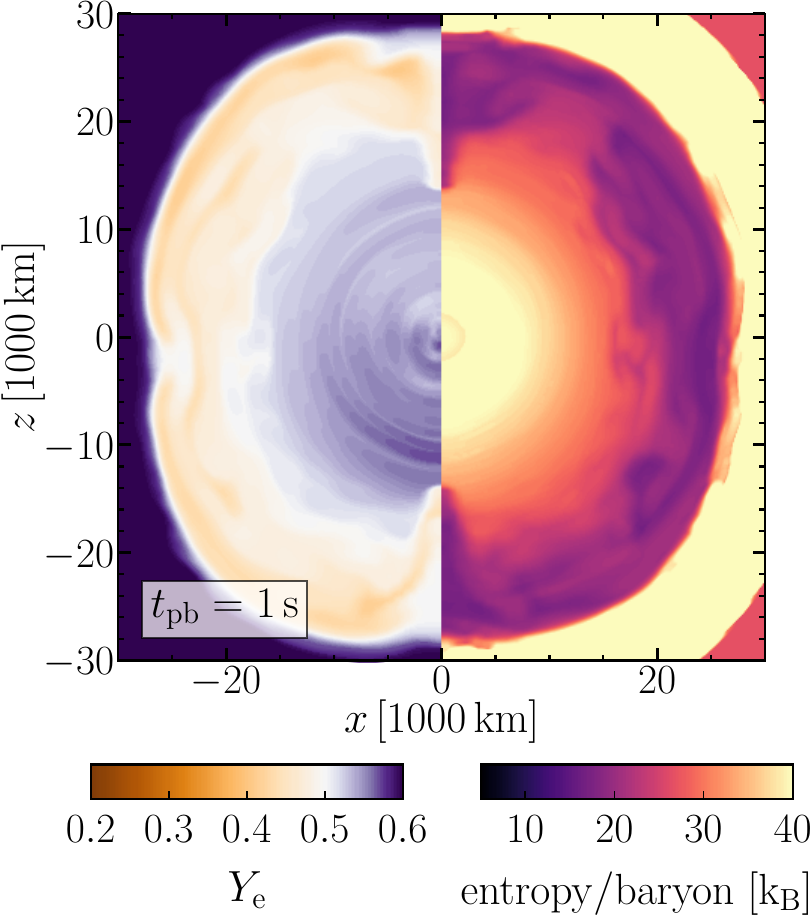}
    \includegraphics[width=0.33\textwidth]{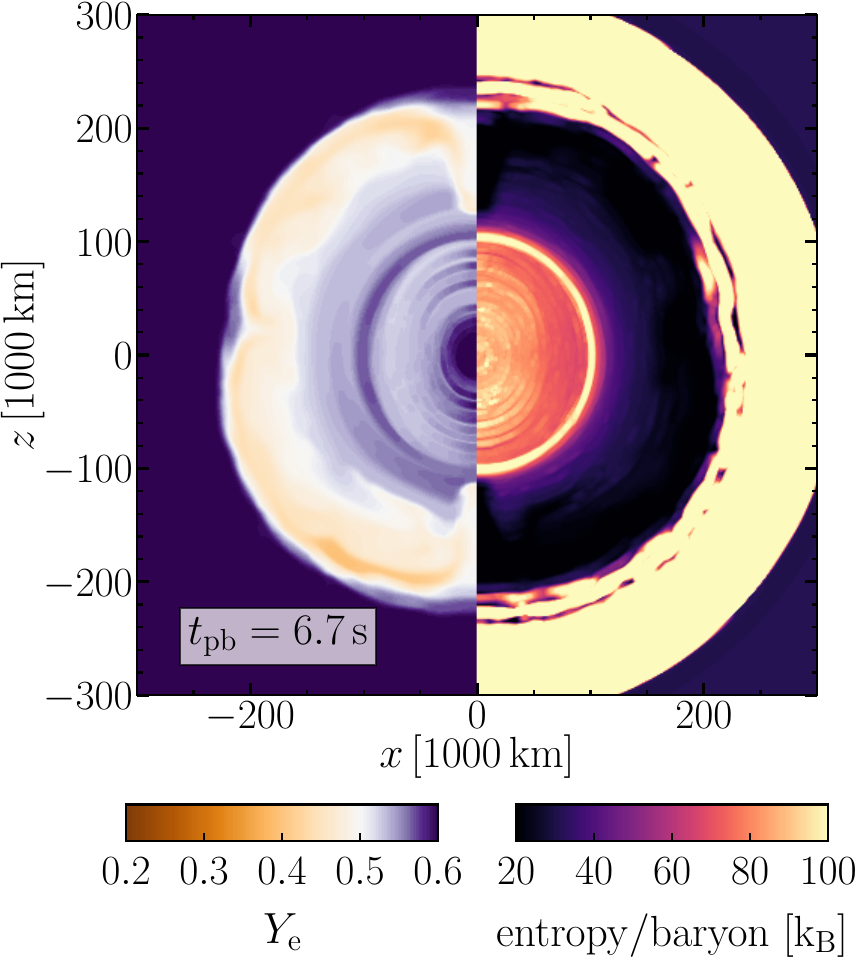}
    \caption{Hydrodynamic evolution of the non-rotating model M1.42-J0 visualized by cross-sectional plots of $\log(\rho)$, radial velocity $v_r$ (left and right half-panels, respectively, in the upper row), $Y_e$, and entropy per baryon (left and right half-panels, respectively, in the lower row) at post-bounce times of 0.1\,s, 1.0\,s, and at 6.7\,s near the end of the simulation (from left to right, as indicated by labels in the plots).}
    \label{fig:M142}
\end{figure*}

\section{Results}
\label{sec:results}

The collapse of all models is triggered by the onset of electron captures. The duration of the infall until core bounce varies between 0.0377\,s and 0.8248\,s (Table~\ref{tab:sims}), depending strongly on the central density and the angular momentum of the initial WD. The bounce time is defined as the first moment when the entropy per nucleon (in units of Boltzmann's constant $k_\mathrm{B}$) at the interface between inner core and supersonically infalling outer core exceeds the value of 3, signaling the formation of a bounce shock. Our lower-density WDs, where the electron captures are quite slow at the beginning, take roughly 10 times longer to reach bounce. This can be concluded from a comparison of the non-rotating case M1.42-J0 with model M1.42-J0.23-Dl, whose rotation has only a weak influence on the collapse of the inner core. In contrast, higher angular momentum stretches the collapse time until bounce by up to a factor $\sim$2.5, especially for the low-density WDs. 

The central density at core bounce correlates inversely with the ratio of rotational to gravitational energy of the inner core, $\beta_\mathrm{ic,b}$ at this moment \citep[Table~\ref{tab:sims}; see also figure~9 in][]{Abdikamalov+2010}. Four of our investigated AIC models (M1.42-J0, M1.42-J0.23-Dl, M1.61-J0.47, and M1.91-J1.09) experience the usual nuclear bounce caused by the stiffening of the EoS when the phase transition to homogeneous nuclear matter takes place near the nuclear saturation density of $\rho_0 \sim 2.6\times 10^{14}$\,g\,cm$^{-3}$. For this subset of models the central density at core bounce also correlates inversely with the value $\beta_\mathrm{i}$ of the initial WD. In the model with the highest angular momentum, M1.91-J1.09, the bounce density reaches only close to $\rho_0$ and drops well below this value right after bounce because of the centrifugal support. Since in this case the collapse does not proceed as far as in the other models and the deleptonization is therefore less strong, we find a significantly higher mass of the inner (i.e., unshocked) core at bounce,\footnote{The inner core at the moment of core bounce is determined as the mass of all matter where the fluid velocity is lower than the local speed of sound at radii $r \le 150$\,km, which turned out to well enclose the shock-formation radius in all directions.} namely $\sim$0.71\,M$_\odot$, whereas in the other three models the inner core masses are between $\sim$0.54\,M$_\odot$ and $\sim$0.59\,M$_\odot$ (Table~\ref{tab:sims}), which is more typical of non-rotating, collapsing stellar cores.

In contrast, the other two AIC models (M1.61-J0.78-Dl and M1.91-J1.63-Dl) undergo a centrifugal bounce at central densities far below $10^{14}$\,g\,cm$^{-3}$ \citep[see also][]{Abdikamalov+2010}. Since these two models with their low initial central densities and big angular momenta contract from great initial radii, their inner cores spin up considerably and carry the largest amounts of angular momentum. We therefore find higher values of the inner core masses at bounce ($\sim$1.03\,M$_\odot$ and $\sim$0.92\,M$_\odot$, respectively; Table~\ref{tab:sims}) and corresponding values of the $\beta$ parameter ($0.123$ and $0.130$, respectively; Table~\ref{tab:sims}). Thus we conclude that a nuclear bounce seems to correlate with $\beta_\mathrm{ic,b}\lesssim 0.1$ and a centrifugal bounce with higher values, but we do not have sufficiently many models to determine the boundary between the two regimes more precisely. The AIC simulations reported by \citet{Abdikamalov+2010} suggested a centrifugal bounce for $\beta_\mathrm{ic,b} \gtrsim 0.2$. This higher threshold than in our models might be connected to the simplified deleptonization treatment during the WD collapse via a $\overline{Y}_{\!e}(\rho)$ trajectory based on transport calculations with a variety of approximations instead of our more elaborate neutrino transport with the \textsc{Alcar} code \citep[see the detailed discussion by][]{Abdikamalov+2010}.

In the following we will discuss our results for the dynamical post-bounce evolution, neutrino emission, and ejecta properties in detail mainly (but not exclusively) for two representative models, namely for the non-rotating case of M1.42-J0, which can be compared to stellar core-collapse simulations of ECSN-like explosions, and for model M1.91-J1.09, which is an example of an AIC of a rapidly and differentially rotating WD with a bounce near nuclear saturation density.

\begin{figure*}
    \centering
    \includegraphics[width=0.32\textwidth]{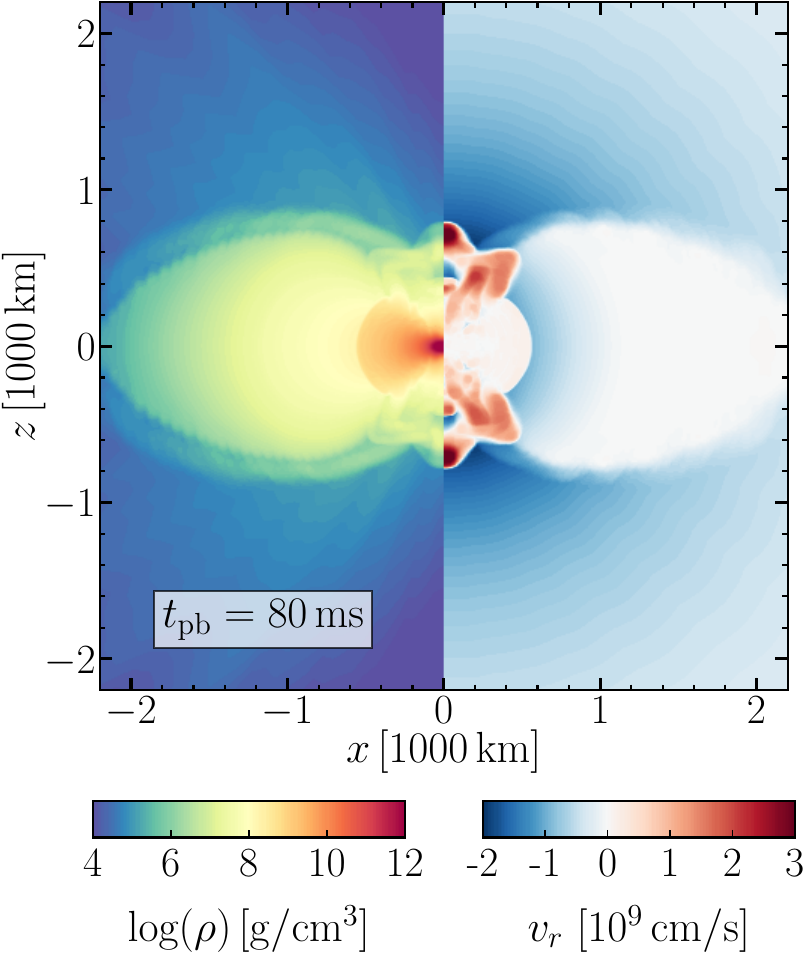}
    \includegraphics[width=0.324\textwidth]{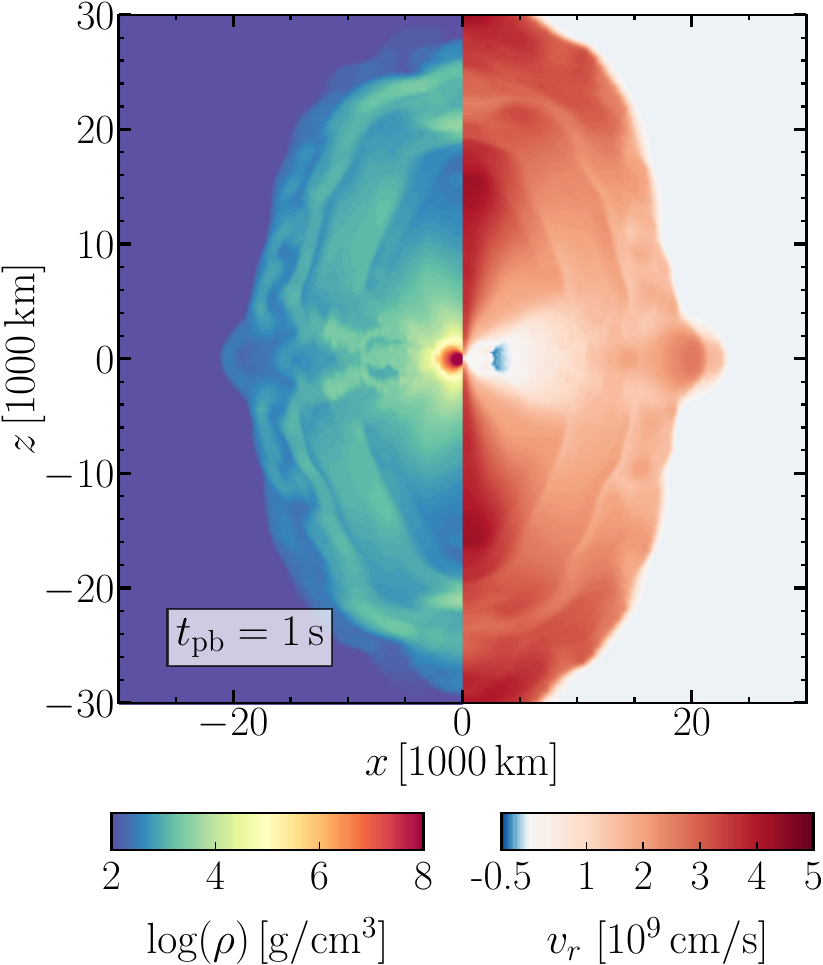}
    \includegraphics[width=0.332\textwidth]{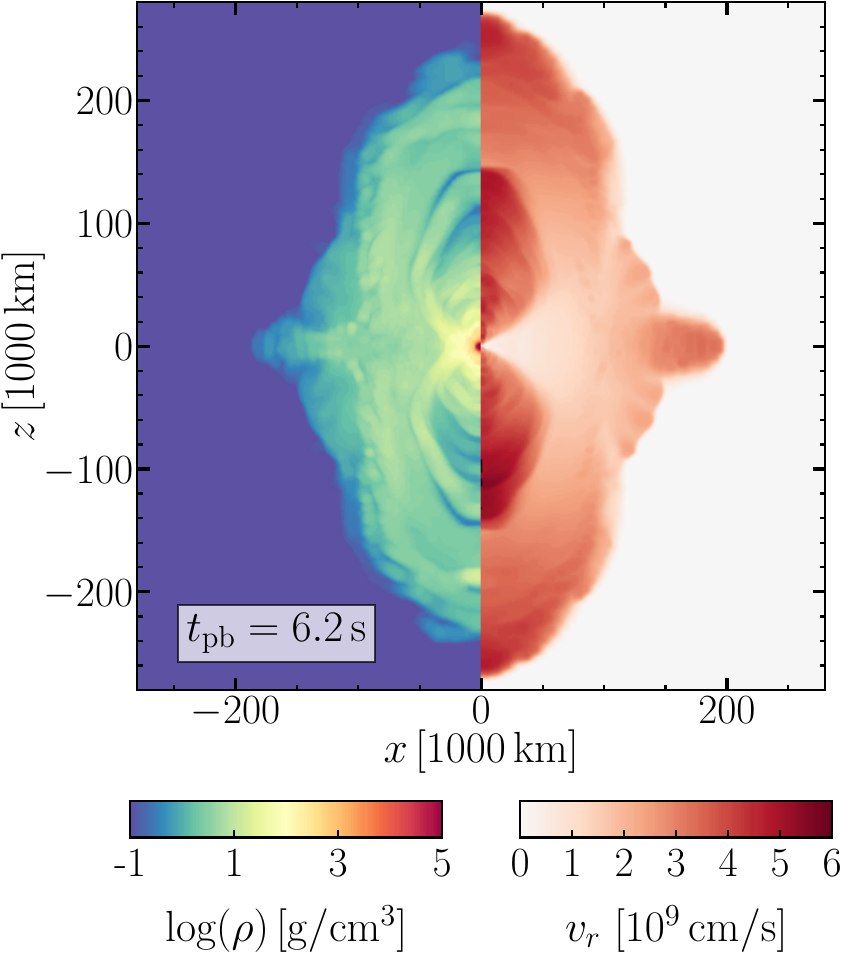}\\\vspace{3pt}
    \includegraphics[width=0.325\textwidth]{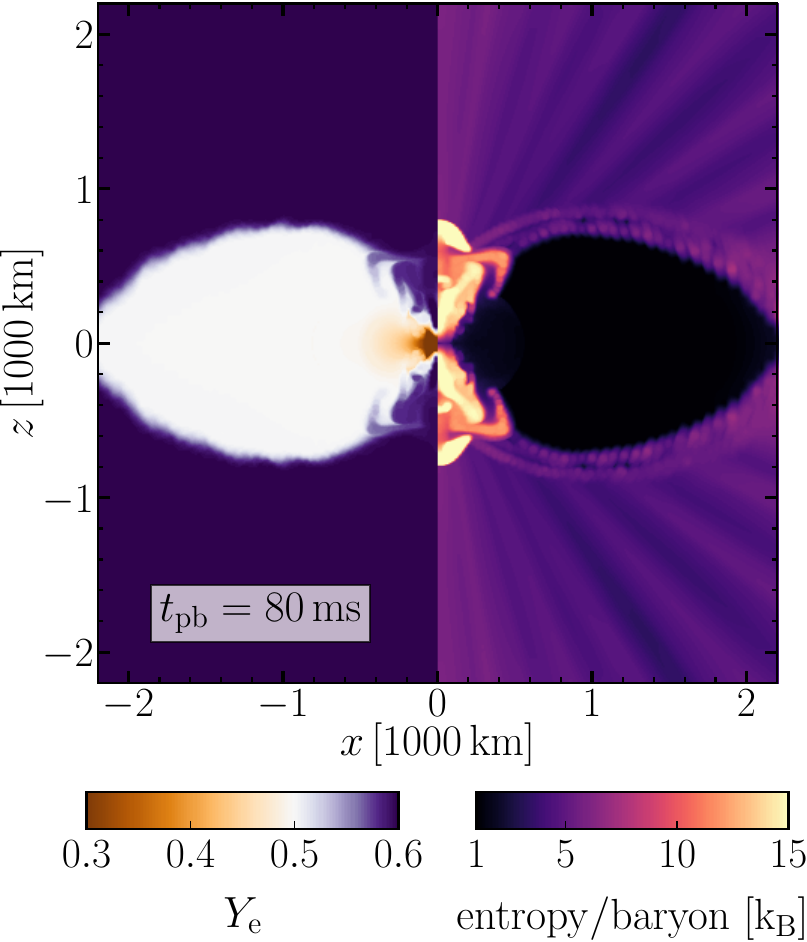}
    \includegraphics[width=0.330\textwidth]{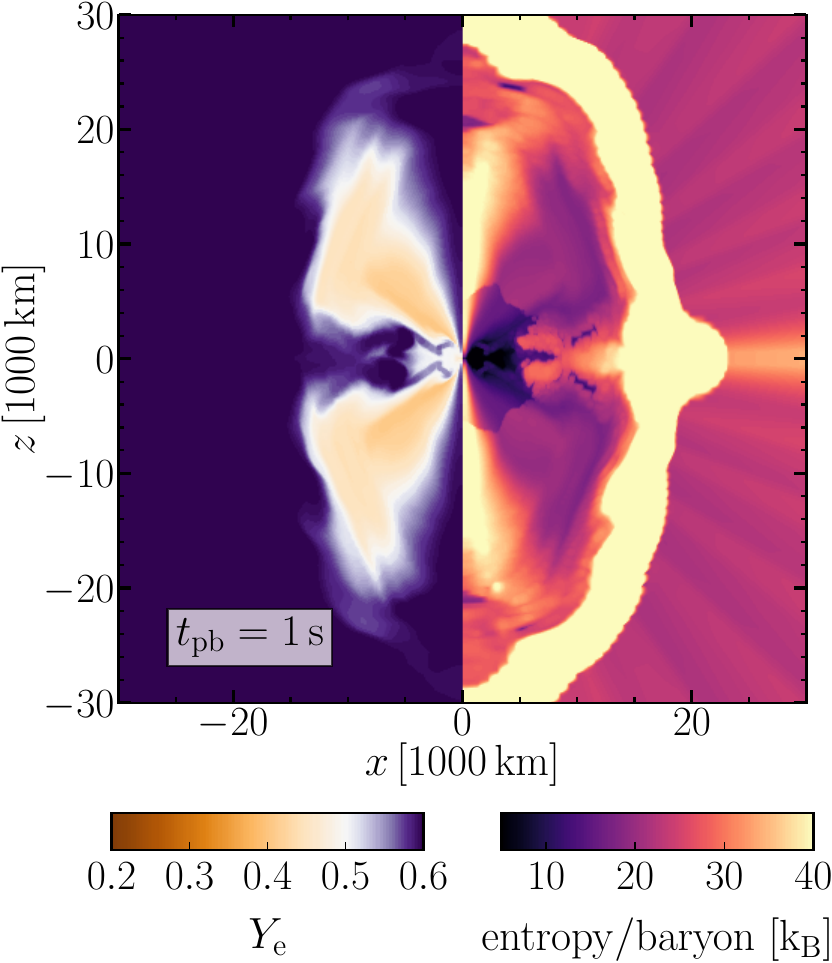}
    \includegraphics[width=0.334\textwidth]{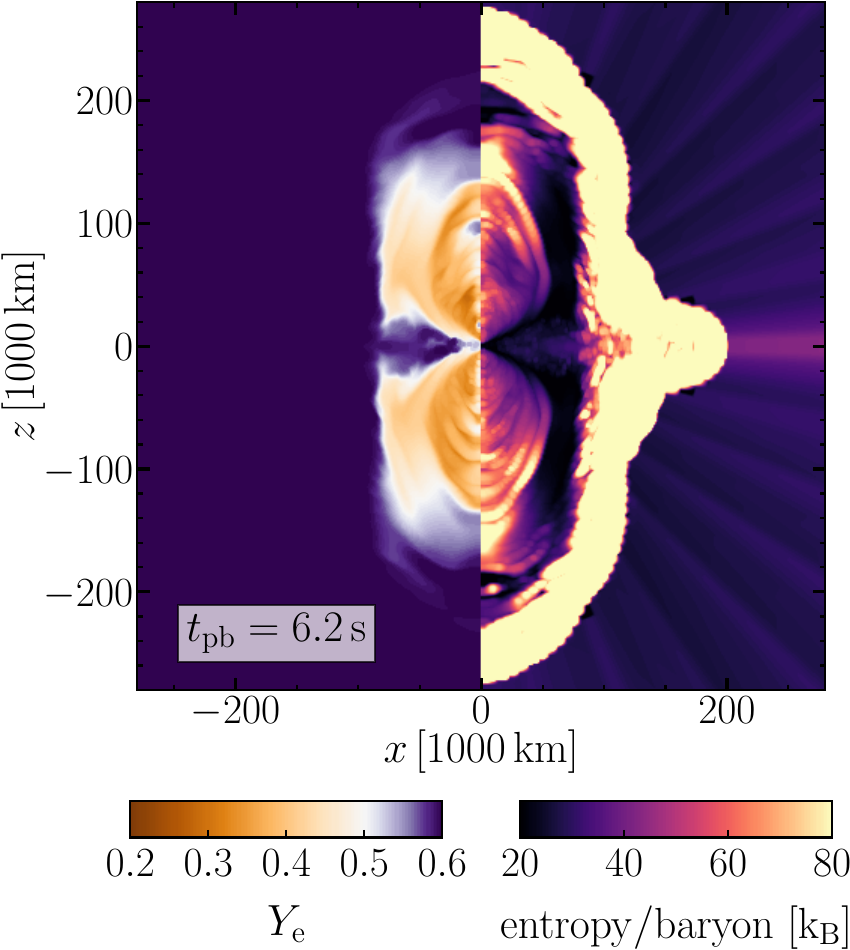}
    \caption{Hydrodynamic evolution of the rapidly rotating model M1.91-J1.09 visualized by cross-sectional plots of $\log(\rho)$, radial velocity $v_r$ (left and right half-panels, respectively, in the upper row), $Y_e$ and entropy per baryon (left and right half-panels, respectively, in the lower row) at post-bounce times of 0.08\,s, 1.0\,s, and at 6.2\,s at the end of the simulation (from left to right, as indicated by labels in the plots).}
    \label{fig:M191-J109}
\end{figure*}

\subsection{Hydrodynamical post-bounce evolution}
\label{sec:hydroevolution}

The energy, mass, and physical properties of matter that is ejected during an AIC are measured for the gas flow that passes a chosen surface (spherical around the non-rotating initial WD and spheroidal around the other, rotating WDs) with positive radial velocity and positive specific total energy (i.e., internal plus gravitational plus kinetic, including rotational energy). Time integration for the ejecta is started between 0.07\,s and 0.10\,s post bounce, at which time the shock breaks out from the collapsing WD and matter begins to be expelled into the surrounding space. Figure~\ref{fig:outflow-mass} displays the time-integrated ejecta energies and corresponding cumulative ejecta masses as functions of time for our six AIC simulations.

The outflow-evaluation surface is placed just above the WD surface (i.e., a few kilometers outside the WD at the equator) such that it envelopes the progenitor as tightly as possible in order to minimize the volume between WD and sphere or spheroid. Nevertheless, we had to post-correct the ejecta masses for the material initially in this volume. As said in Section~\ref{sec:progenitors}, our CSM density close to the WD surface should have been chosen lower to reduce the amount of this artificial background medium and to make sure that the CSM mass is far dominated by the true WD ejecta. Moreover, due to the initial limit of the EoS table (which was eliminated only later), we had set $Y_e$ in the CSM to 0.6, which complicated the discrimination of the true ejecta from the CSM in our rotating models, where $Y_e$ of some ejected matter reaches beyond~0.6.

\subsubsection{Non-rotating model M1.42-J0}

The hydrodynamical evolution of the non-rotating model M1.42-J0 displays the well-known features from core-collapse and explosion simulations of ECSNe and ECSN-like cases in 2D and 3D \citep{Janka+2008,Wanajo+2011,Janka+2012,Melson+2015,Radice+2017,Stockinger+2020,Wang+2024}. The SN shock formed at core bounce expands continuously and as it runs down the steep density gradient at the edge of the WD or degenerate stellar core, its outward motion accelerates to very high velocities of several 10,000 km\,s$^{-1}$ within less than 100\,ms after bounce. Neutrino heating in the gain layer between gain radius and shock front supplies the blast wave with energy and determines the explosion energy of the ejecta. 

While the shock remains essentially spherical, the neutrino energy deposition causes a short phase of convective activity in the gain layer, in course of which Rayleigh-Taylor mushrooms create asymmetries in the earliest neutrino-heated ejecta. The fast, buoyancy-driven rise of these structures permits initially neutron-rich matter to retain relatively low $Y_e$ values down to less than 0.4 (Figure~\ref{fig:M142}), in contrast to 1D simulations, where the $\nu_e$ and $\bar\nu_e$ interactions in the earliest ejecta navigate $Y_e$ to a more narrow range around 0.5 \citep[see][]{Wanajo+2009,Wanajo+2011}. When the convectively perturbed, earliest ejecta shell moves outward behind the shock, it is followed by a nearly spherical baryonic wind blown off the PNS surface due to the energy transfer by neutrinos to matter just outside the neutrinospheres. As time advances, the electron fraction and entropy per nucleon in this neutrino-driven wind increase (up to about 0.6 and 80--100\,$k_\mathrm{B}$, respectively) and the wind fills the growing volume between the inhomogeneous postshock shell at larger radii and the PNS at the center (Figure~\ref{fig:M142}).

For model M1.42-J0 the energy of the ejecta (``explosion energy,'' $E_\mathrm{ej}$, in Table~\ref{tab:sims}) asymptotes to about $1.1\times 10^{50}$\,erg. This result is in good agreement with the energies of (non-rotating) ECSN and ECSN-like explosions published by the Garching group \citep[e.g.,][]{Kitaura+2006,Janka+2008,Melson+2015,Stockinger+2020}. Minor, case-dependent differences are caused by the fact that the SN simulations there were started from stellar evolution models that had more realistic ONe and low-mass Fe cores with slightly smaller masses, lower initial $Y_e$, and differences in the nuclear composition, density, and temperature compared to the WDs considered here. Moreover the simulations were performed with different nuclear EoSs and different codes (\textsc{Vertex} for the SN models instead of \textsc{Alcar} here) using different electron-capture rates on nuclei during core collapse. Such differences led to higher binding energies of the initial stellar cores and to inner (unshocked) cores at bounce that were smaller by roughly 0.1\,M$_\odot$ in the models of ECSN and ECSN-like explosions. 

Note that the values for $E_\mathrm{ej}$ listed in Table~\ref{tab:sims} are the total ejecta energies computed as the sums of thermal plus degeneracy, kinetic, and gravitational energies. For this summation we have corrected (reduced) the internal energies evolved by the hydro code for the rest-mass contributions of electrons, nucleons, and nuclei according to the local composition at the outflow-evaluation surface,\footnote{Like all nuclear EoSs for the NSE regime, the energy density provided by the SFHo EoS is normalized to a reference value of the average nucleon mass and includes the remaining rest-mass energies of nucleons and nuclei as well as rest-mass energies of electrons and positions. The true internal energy, i.e., the thermal and degeneracy energy of the particles responsible for the gas pressure, therefore requires subtracting the rest-mass contributions of all particles for a given (local) nuclear composition. Because of this dependence on the composition, which accounts for energy conversion in nuclear reactions, the values of the total energy can vary with the location of measurement due to changes of the gas composition in space and time.} and we have checked that these energies will not excessively change due to subsequent changes of the nuclear composition in the expanding matter. In this context, we provide two numbers for the ejecta energies in Table~\ref{tab:sims}, one that is measured for the mass streaming through the outflow-evaluation surface with the composition at this point, and a second value in parentheses obtained under the assumption that all ejecta end up in a representative iron-group nucleus. These two numbers provide a rough insight into the sensitivity of $E_\mathrm{ej}$ to the nuclear composition, but they do not necessarily bracket the true values, since the accuracy of such estimates is limited by the NSE approximation used in our simulations. Moreover, the later energy release by the decay of radioactive isotopes (which are not considered in our treatment) may have an impact on the long-time evolution of the outflow, e.g., it may cause a significant boost of the expansion velocities.

\subsubsection{Rapidly rotating model M1.91-J1.09}
\label{sec:hydro-rap-rot-mod}

Model M1.91-J1.09 has the largest initial angular velocity (w.r.t.\ the central value as well as the maximum value) among all of our models. Nonetheless, it still reaches a central density at core bounce close to the nuclear saturation value, owing to its high initial density. In contrast, models M1.61-J0.78-Dl and M1.91-J1.63-Dl experience their centrifugal bounce at much lower central densities. Also the post-bounce dynamics of all of our rotating models differs fundamentally from that of the non-rotating model.

Because of the faster infall of gas around the poles compared to matter near the equator, a stronger bounce shock forms at the poles in all of these models, causing wide-angle outflows preferentially towards the polar directions. Close to the equator, centrifugal hang-up leads to a much weaker shock, which converts to a stalled accretion shock within 100--200\,ms, depending on the model, thus not causing any mass ejection. Instead, an extended, torus-like equatorial bulge of centrifugally supported WD matter assembles around the newly forming PNS (left panels of Figure~\ref{fig:M191-J109}),\footnote{In the lower panels of Figure~\ref{fig:M191-J109} a stripe-like pattern of regular entropy variations with polar angle $\theta$ can be seen in the assumed CSM. These variations correlate with density variations, which are not visible in the plots because of the color saturation at the lowest CSM densities. The density (and consequential temperature and entropy) variations are caused by our method to compute the initial WD models on a polar grid. In the case of WDs that are rotationally deformed, the surface is not aligned with a radial coordinate line and therefore the minimum density reached at the surface is not uniform but varies up and down with some periodicity (but without a systematic trend) as a function of polar angle $\theta$. These variable WD surface densities were then used as base points for the CSM constructed according to the equations mentioned in Section~\ref{sec:progenitors}. By this procedure the surface-density variations of the WD models were mapped into the CSM. The low-level variations in density, temperature, and entropy did not significantly affect our hydrodynamic results beyond the effects anyway caused by the too high CSM densities, as mentioned in Sections~\ref{sec:progenitors} and~\ref{sec:hydroevolution} and several other places. Moreover, since all reported hydrodynamic outflow properties are monitored for the ejecta flowing through a spheroidal surface that is placed very close to the initial WD surface (see Section~\ref{sec:hydroevolution}), the CSM and its properties has effectively only a marginal influence on the results discussed in our paper, because the spheroid is passed by the first ejecta soon after they are launched and the subsequent outflow is not in touch with the CSM.} whose mass slowly increases by continuous accretion from the equatorial bulge. Table~\ref{tab:sims} provides the PNS and torus masses at the end of our simulations. We define the PNS's mass here by the criterion that its matter has densities higher than $10^{10}$\,g\,cm$^{-3}$ instead of $10^{11}$\,g\,cm$^{-3}$ used in previous papers for PNSs formed in stellar core collapse events. This accounts for the fact that the rapidly spinning PNSs emerging from our AIC models posses centrifugally inflated mantle regions of fairly low average densities, whereas for non-rotating PNSs the different density criteria imply only tiny mass differences at late times when the density gradient at the PNS surface becomes very steep. The extended, equatorial, rotationally stabilized and neutrino-transparent accretion torus, which surrounds the oblately deformed high-density PNS core with its overlying, bulge-shaped mantle region in our rotating AIC models, is defined by material that has densities $\rho \le 10^{10}$\,g\,cm$^{-3}$ and is gravitationally bound (with total specific energies less than or equal to zero). We also apply the criterion that $Y_e \le 0.55$ in order to avoid the inclusion of any admixture from the assumed CSM in our budgeting of the torus mass.  

Simultaneously to feeding the PNS, the equatorial accretion also adds more material into the polar outflows, since some fraction of the matter flowing inward mainly along the torus surface gets neutrino heated near the PNS and is expelled again into the polar funnels. The shock front enveloping the polar ejecta spreads around the equatorial torus within a few 100\,ms (middle panels of Figure~\ref{fig:M191-J109}), in course of which the initially collimated ejecta also spread sideways and the polar outflow funnels widen. Analogue to the situation in the non-rotating model, the shock detaches from the ejecta cloud as it speeds through the low-density CSM material. The final half-opening angle of the hour-glass shaped polar outflows in model M1.91-J1.09 is 70$^\circ$--80$^\circ$ at large distances (right panels of Figure~\ref{fig:M191-J109}). In other models the outer contours of the outflows more closely resemble a butterfly, and the half-opening angles at the end of our simulations vary between $\sim$45$^\circ$ and $\sim$85$^\circ$ far away from the PNS. The outflows are stratified in the latitudinal direction with proton-rich cores around the polar axis and more neutron rich flows at larger angles from the polar axis. The cores become less proton-rich and thinner as time progresses and the early bipolar ejecta with proton excess are followed by dominantly neutron-rich matter in the whole outflow cones at late times.

Figure~\ref{fig:outflow-mass} reveals a steepening of the growth of the cumulative ejecta masses $M_\mathrm{ej}(t)$ and ejecta energies $E_\mathrm{ej}(t)$ at $t_\mathrm{pb} > 3$\,s in all of the rotating models except M1.42-J0.23-Dl. This late-time effect is connected to the onset of episodic phases of enhanced inflow of torus gas, which happens predominantly in the surface layers of the thick tori, and subsequent eruptive re-ejection of neutrino-heated material, thus amplifying the net mass-outflow rates by roughly a factor of 10 (from typically a few $10^{-4}$\,M$_\odot$\,s$^{-1}$ before the episodic variations begin to some $10^{-3}$\,M$_\odot$\,s$^{-1}$ time-averaged afterwards). Such unsteady flow dynamics manifests itself in large-amplitude variations of the density and radial velocity in the polar outflows, with velocities alternating between positive and negative values close to the base of the mass ejection. During these later evolutionary stages, secondary shocks drive bubbles and plumes of matter with high velocities and increasing entropies outward along the rotation axis (see the fine structures in the polar outflow lobes in the right panels in Figure~\ref{fig:M191-J109}). The temporal variations are caused by inflowing and outflowing matter interacting in unstable shear layers that interface the near-equator regions of the accretion tori and the polar ejecta in the close vicinity of the PNS. As a consequence, either due to the development of Kelvin-Helmholtz instability or the decreasing outward push by neutrino heating at later times, the inflow at the torus surface becomes unsteady and triggers temporal and thus radial variability in velocity, density, $Y_e$, and entropy of the outflows as well.

Model M1.42-J0.23-Dl does not display the steeper rise of $M_\mathrm{ej}(t)$ and $E_\mathrm{ej}(t)$ at $t_\mathrm{pb} > 3$\,s, because its higher neutrino luminosities and mean energies permit more intense neutrino heating and stronger, continuously fast polar mass ejection even at late times. Such conditions seem to counteract the development of the instabilities in the shear layer between the outflows and the torus. Model M1.42-J0.23-Dl reaches the onset of the unsteady inflow-outflow behavior only shortly before our simulation was stopped.

The detailed flow dynamics in the shear layer and its consequences for PNS accretion, neutrino emission, and outflow heating are resolution dependent to some extent, but the resolution tests that we performed for comparison revealed that the overall evolution of the AICs including their mass-outflow rates and cumulative ejecta masses and energies do not change because of small-scale differences between our standard-resolution simulations and the higher-resolution runs. An analysis of the described hydrodynamical phenomena in more depth is postponed to a follow-up paper.

\begin{figure*}
    \centering
    \includegraphics[width=\textwidth]{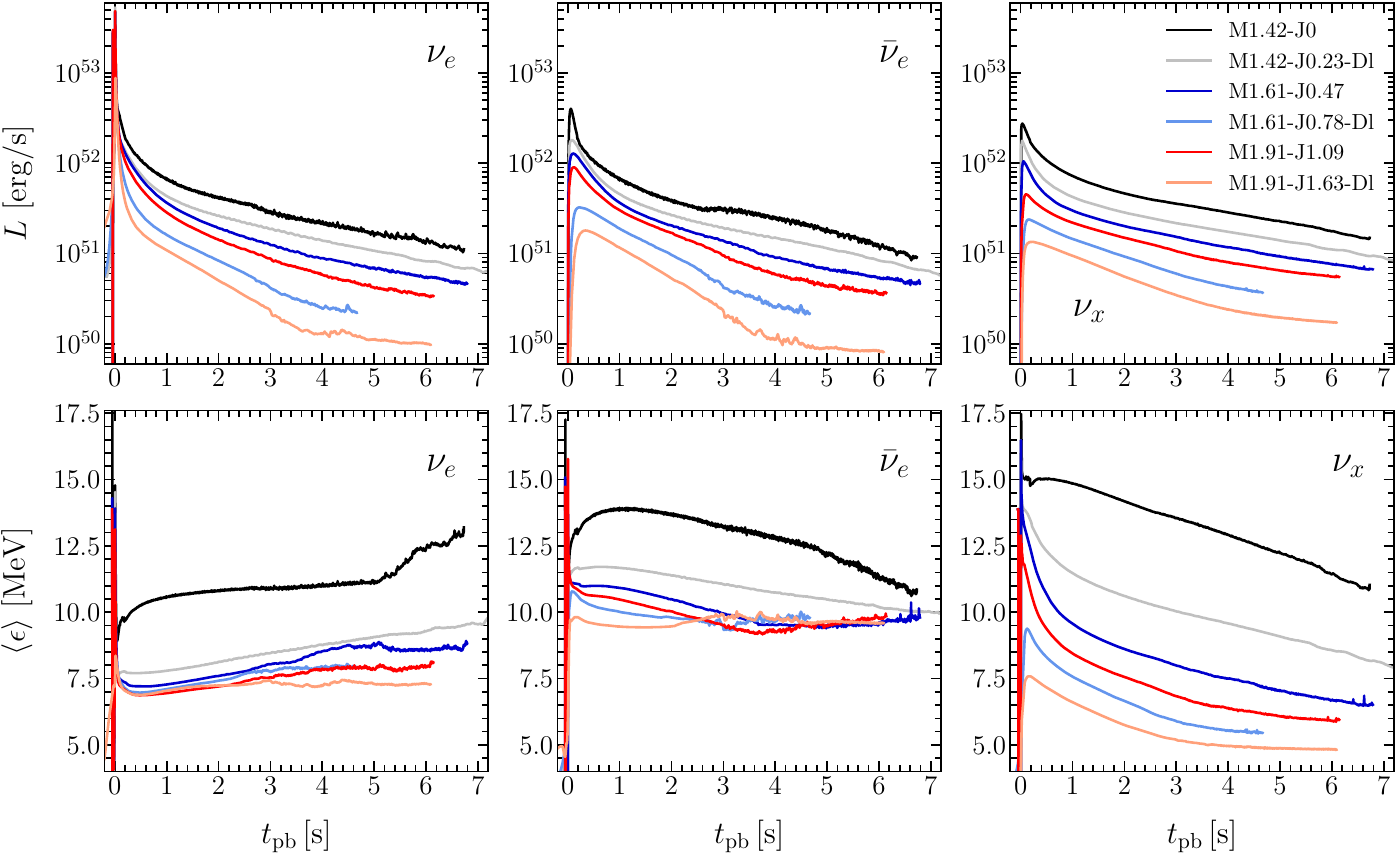}
    \caption{Radiated neutrino luminosities $L$ ($4\pi$-integrated total energy fluxes) and direction-averaged mean energies $\langle \epsilon\rangle$ (luminosities divided by direction-integrated number flux) versus time normalized to core bounce for all of our AIC simulations, measured in the observer frame at a radius of 3500\,km, and time-shifted by $-10$\,ms to (approximately) account for the light travel time from the neutrinospheric emission region to 3500\,km. The left panels show the results for $\nu_e$, the middle panels those for $\bar\nu_e$, and the right panels those for a single species of $\nu_x$. The steep increase of the mean energy of $\nu_e$ in model M1.42-J0 at $t_\mathrm{pb} \gtrsim 5$\,s is a numerical artifact (see Section~\ref{sec:nuemission} for details).}
    \label{fig:lum-meanE}
\end{figure*}

\begin{figure*}
    \centering
    \includegraphics[width=0.325\textwidth]{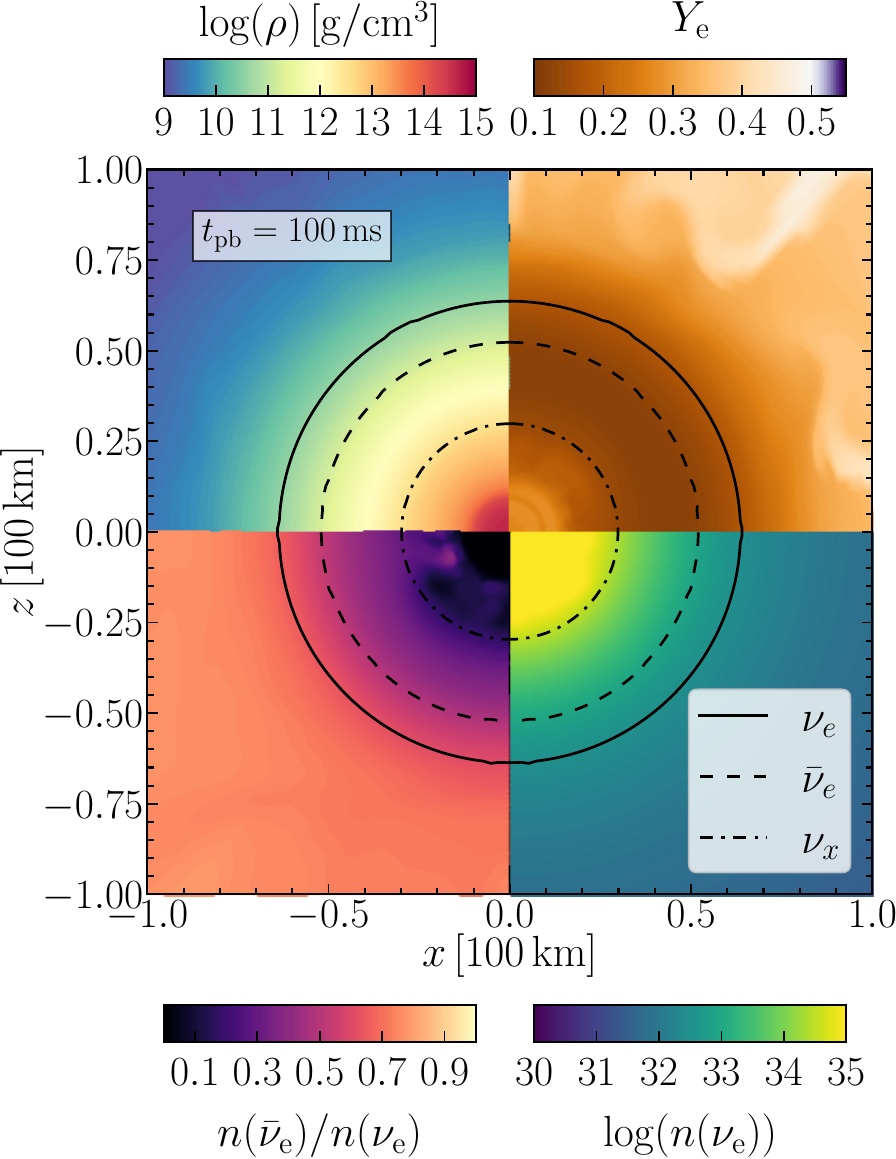}
    \includegraphics[width=0.325\textwidth]{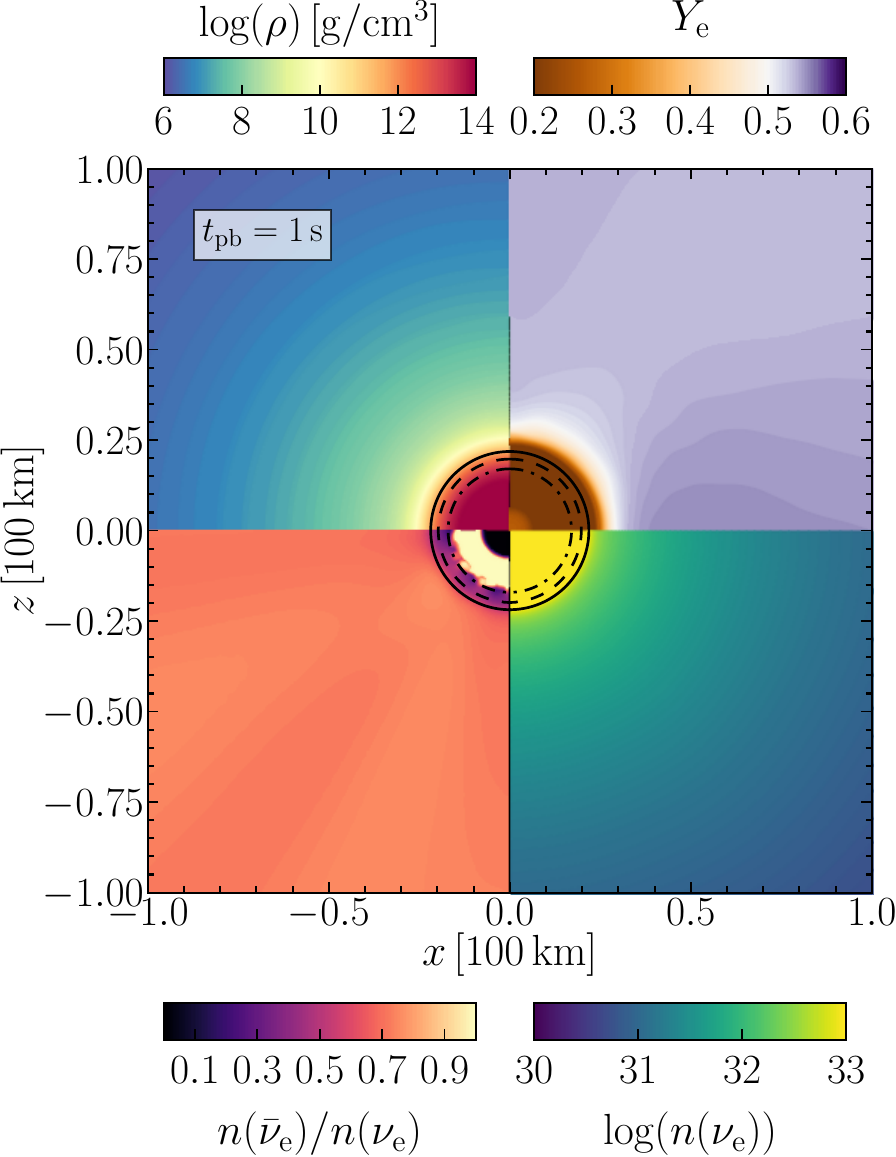}
    \includegraphics[width=0.325\textwidth]{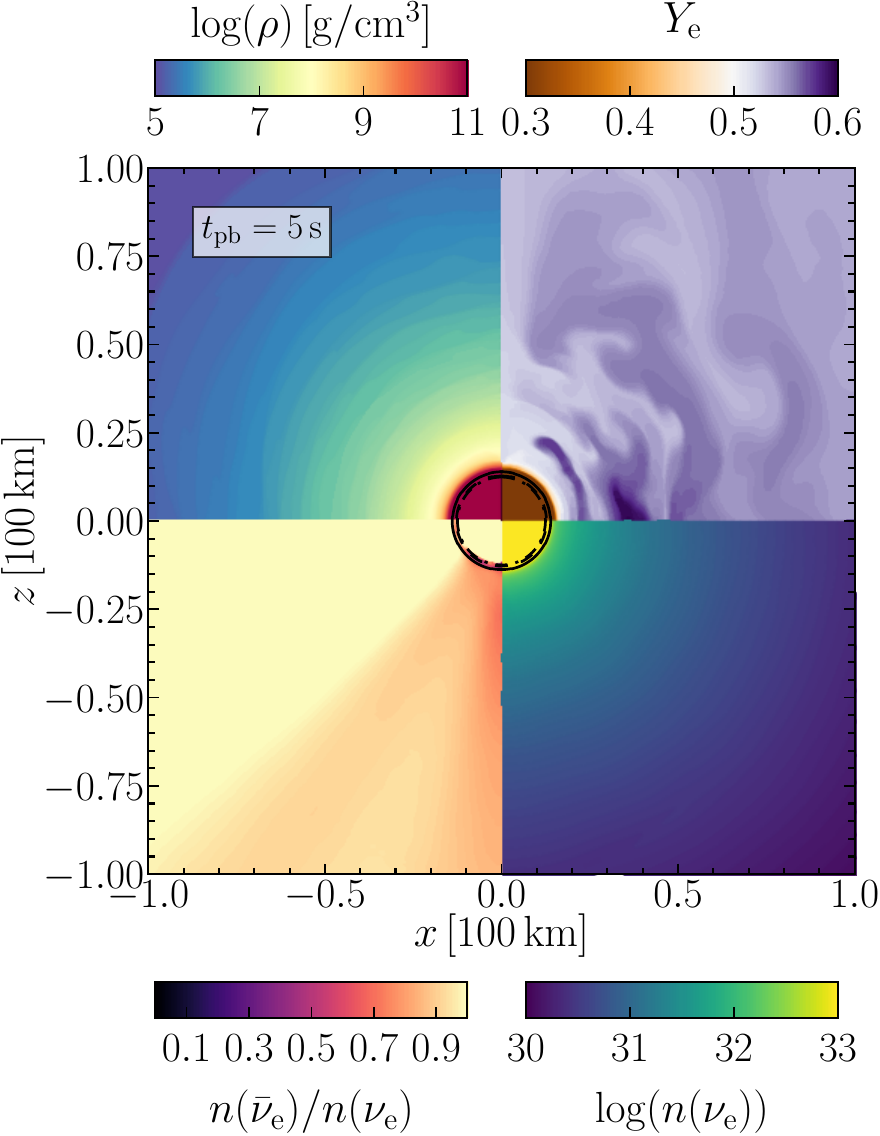}
    \caption{Neutrinospheric region and close surroundings of our non-rotating model M1.42-J0 at 0.10\,s, 1.00\,s and 5.00\,s near the end of our simulation. Only one hemisphere of the full-180$^\circ$ 2D calculation is displayed. The solid, dashed, and dash-dotted circular lines mark the average neutrinospheres (energy-averaged over the spectrum of the radial energy fluxes) of $\nu_e$, $\bar\nu_e$, and $\nu_x$, respectively. The quarter-panels, clock-wise from top left, show $\log(\rho)$, $Y_e$, the logarithm of the electron-neutrino number density, $\log(n(\nu_e))$, and the ratio of $\bar\nu_e$ to $\nu_e$ number densities. The corresponding color bars are provided above and below the three panels. Note that the ranges of values for $\log(\rho)$, $Y_e$, and $\log(n(\nu_e))$ differ between the three panels in order to better show the variations exterior to the neutrinospheres. The (fluctuating) nonspherical perturbations in the surroundings of the PNS are caused by postshock convection (at 100\,ms p.b.) and low-level neutrino-emission asymmetries due to PNS convection (at 1\,s and 5\,s after bounce).}
    \label{fig:M142-dens-neutrinospheres}
\end{figure*}

\subsection{Neutrino emission}
\label{sec:nuemission}

Our non-rotating model M1.42-J0 exhibits properties of the neutrino emission (i.e., luminosities and mean energies of all neutrino species as functions of time; see Figure~\ref{fig:lum-meanE}) and of the PNS evolution (Figure~\ref{fig:M142-dens-neutrinospheres}) that are fully compatible with the ECSN-like case of model z9.6 in \citet{Mirizzi+2016} and with the neutrino-cooling simulation of the lowest-mass PNS model in \citet{Fiorillo+2023}. In both cases also the SFHo EoS was used and the PNS had a mass of about 1.36\,M$_\odot$, i.e., quite similar to the 1.41\,M$_\odot$ of our non-rotating AIC model. Moreover, the PNS was born in a low-density mantle of the progenitor, which had no dynamically relevant influence on the explosion and permitted an explosion after only a short post-bounce accretion phase. The simulations in \citet{Mirizzi+2016} and \citet{Fiorillo+2023} were performed in 1D but took into account PNS convection by a mixing-length treatment, and they employed the \textsc{Vertex} neutrino transport code, which is independent of the \textsc{Alcar} code applied in the present work, since it is based on a different solver for the coupled neutrino moment equations as well as an independent implementation of the neutrino interaction rates. 

Just as one example of the assuring agreement between the \textsc{Vertex} and \textsc{Alcar} results for the neutrino luminosities, we mention the slight kink of the $\nu_e$ and $\bar\nu_e$ luminosities in model M1.42-J0 at $\sim$2.7\,s after bounce, where the decay of the $\nu_e$ luminosity becomes a bit steeper and the decline of the $\bar\nu_e$ luminosity becomes a bit flatter. This feature can also be seen in Figure~12 of \citet{Mirizzi+2016} and correlates with the time when the PNS convection reaches the center of the PNS, thus modifying the subsequent deleptonization history of the PNS. 

However, we also point out some unphysical behavior of the mean $\nu_e$ energy in our model M1.42-J0 connected to shortcomings in the present simulations with \textsc{Alcar}. The mean electron-neutrino energy increases continuously with time instead of showing a local maximum similar to the mean energies of $\bar\nu_e$ and $\nu_x$ at 1.0--1.5\,s after bounce. Moreover, one can witness a steepening of the increase of $\langle\epsilon_{\nu_e}\rangle$ after about 5\,s. The former effect could be traced back to the lack of the single-particle mean-field potentials of neutrons and protons in the charged-current interactions of $\nu_e$ and $\bar\nu_e$ with these nucleons \citep{Reddy+1998,Roberts+2012,Martinez-Pinedo+2012} in our version of the \textsc{Alcar} code, whereas a test calculation with a finer radial grid at $t_\mathrm{pb} > 5$\,s showed that the latter effect is connected to insufficient radial resolution in the neutrinospheric region of $\nu_e$ at late times, when the density decline at the PNS surface becomes extremely cliffy and the radial grid applied in the current model did not track this gradient perfectly well. We will improve future simulations with respect to both aspects.\footnote{We stress that the previously published results from calculations with the \textsc{vertex} code, which we refer to for comparison, do {\em not} suffer from both of these problems and their underlying reasons.} As for the models presented here, we have made sure also by test calculations that these unphysical features in $\langle\epsilon_{\nu_e}\rangle$ do not seriously affect our rotating models, which are most relevant for the present paper, and, in particular, that our main messages remain valid.

\begin{figure*}
    \centering
    \includegraphics[width=0.325\textwidth]{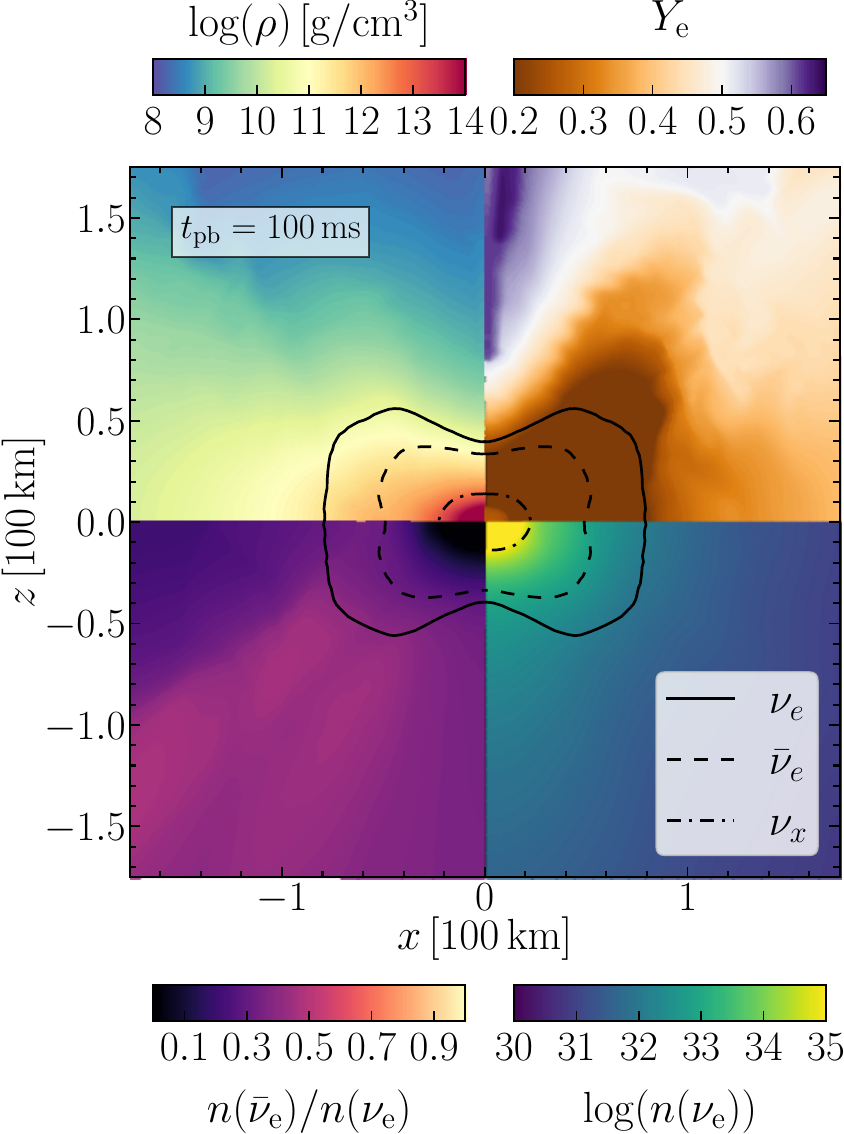}
    \includegraphics[width=0.325\textwidth]{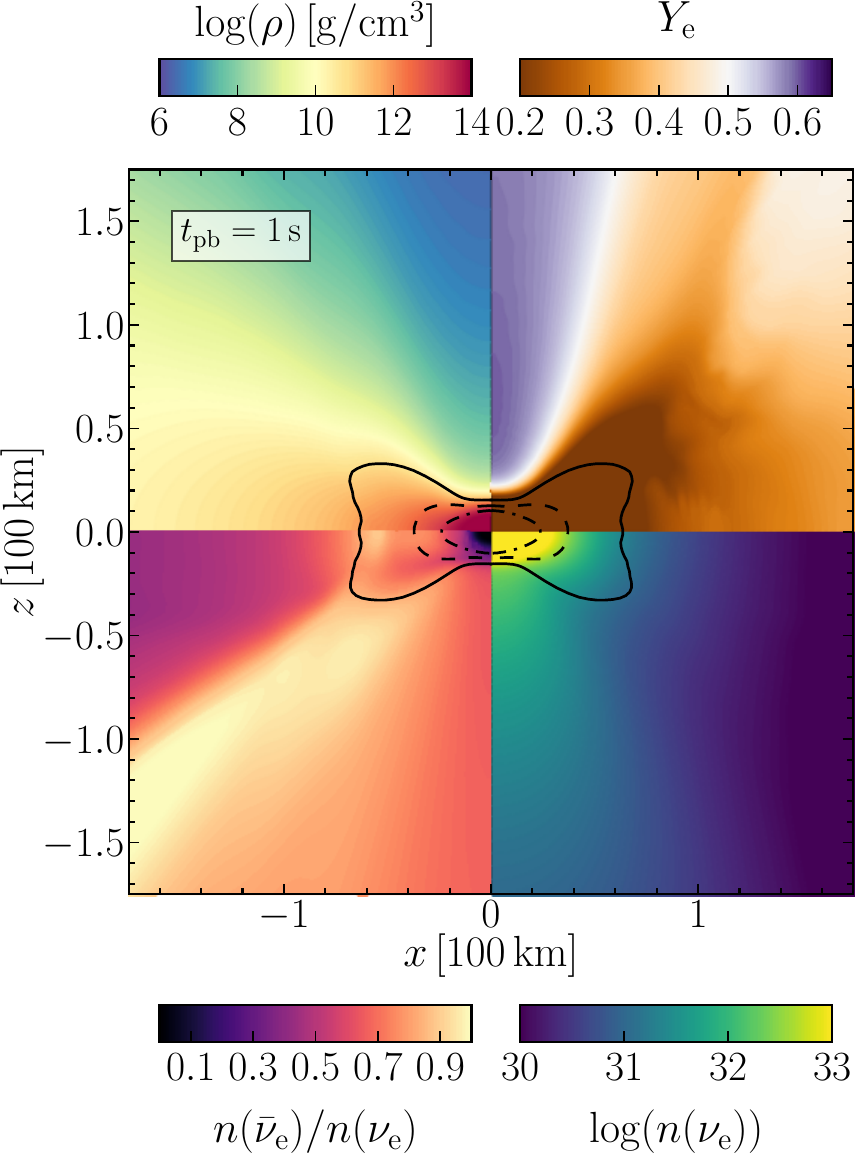}
    \includegraphics[width=0.325\textwidth]{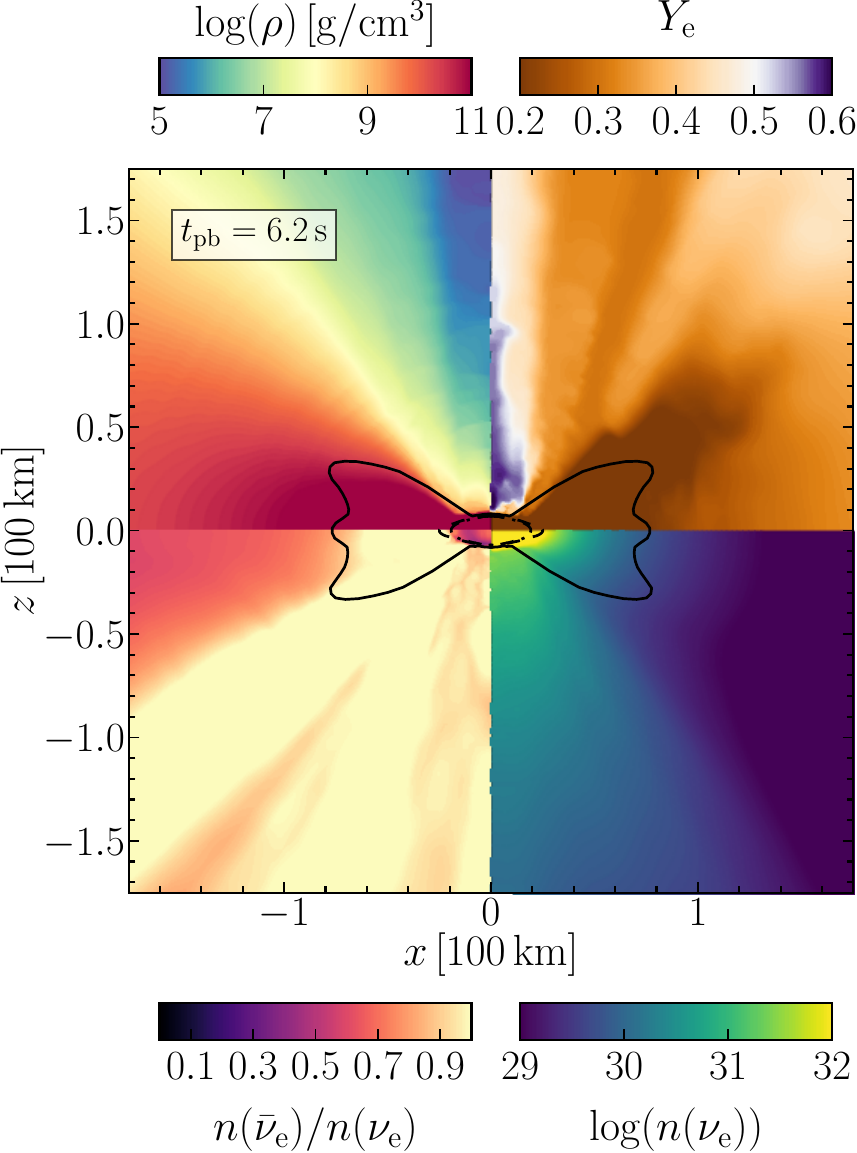}
    \caption{Neutrinospheric region and close surroundings of the rapidly rotating model M1.91-J1.09 at 0.10\,s, 1.00\,s and 6.15\,s at the end of our simulation. Only one hemisphere of the full-180$^\circ$ 2D calculation is displayed. The solid, dashed, and dash-dotted lines mark the average neutrinospheres (energy-averaged over the spectrum of the radial energy fluxes) of $\nu_e$, $\bar\nu_e$, and $\nu_x$, respectively. The quarter-panels, clock-wise from top left, show $\log(\rho)$, $Y_e$, the logarithm of the electron-neutrino number density, $\log(n(\nu_e))$, and the ratio of $\bar\nu_e$ to $\nu_e$ number densities. The corresponding color bars are provided above and below the three panels. Note that the ranges of values for $\log(\rho)$, $Y_e$, and $\log(n(\nu_e))$ differ between the three panels in order to better show the variations exterior to the neutrinospheres.}
    \label{fig:M191-dens-neutrinospheres}
\end{figure*}

Since major contributions of the $\nu_e$ emission in our rotating models and also of the $\bar\nu_e$ emission for most of the simulated evolution periods stem from the centrifugally flattened PNS mantle at $\rho\lesssim 10^{12}$\,g\,cm$^{-3}$ and from the extended, neutrino-transparent accretion tori with densities below $\sim$$10^{10}$\,g\,cm$^{-3}$ (see Figure~\ref{fig:M191-dens-neutrinospheres}), the mean-field potentials in the neutrino absorption and production reactions play a much less important role in the rotating models than in the non-rotating case. Moreover, the density gradients in most of the neutrinospheric regions of the rotating models are more shallow. In this context we tested the impact of higher radial resolution by making use of our rezoning procedure implemented in \textsc{Alcar} and repeating parts of the late evolution phases of some of the rotating models. We witnessed local changes only around the poles near the PNS surface, in particular in the shear interfaces between bipolar outflows and accretion torus, which could be better resolved with the finer radial grid. But there were no significant and important differences in the overall evolution and ejecta properties.

The neutrino luminosities of all neutrino species ($\nu_e$, $\bar\nu_e$, and $\nu_x$) as functions of time display a strict inverse ordering with the ratio of rotational to gravitational energy of the inner core at bounce, $\beta_\mathrm{ic,b}$ (Figure~\ref{fig:lum-meanE}, upper panels, and Table~\ref{tab:sims}), because faster rotating PNSs with accretion tori are, overall, less dense and hot.\footnote{The ${\cal T}/|W|$ ratios of the bound remnants, $\beta_\mathrm{rem,f}$ for the PNS plus torus (Table~\ref{tab:sims}) and $\beta_\mathrm{pns,f}$ for the PNS (the latter are only a few percent smaller than the former) at the final times of our simulations, can also roughly serve as sorting parameters. However, this does not work perfectly well for M1.61-J0.78-Dl and M1.91-J1.63-Dl, whose $\beta$'s for the remnants are reversed compared to those of the inner cores at bounce, because both models possess extremely extended tori and very similar $\beta$ values.} Thus faster rotation clearly has a more important impact on the magnitude of the neutrino luminosities than the mass of the WD, because AIC models with higher angular momentum possess lower neutrino luminosities even when their initial WDs are more massive. Models M1.61-J0.78-Dl and M1.91-J1.63-Dl in comparison to M1.61-J0.47 and M1.91-J1.09, respectively, clearly demonstrate the crucial influence of $\beta_\mathrm{ic,b}$. Despite their identical WD masses and similar initial values $\beta_\mathrm{i}$, the low-density WD models reach considerably higher values of $\beta_\mathrm{ic,b}$. Since their infall starts at substantially larger average radii, angular momentum conservation implies a greater increase of the $\beta$-ratio during collapse: $\beta_2/\beta_1 \propto R_1/R_2$ for a (self-similar) contraction from initial state~1 with representative radius $R_1$ to final state~2 with representative radius $R_2$. Therefore the influence of rotation on the long-term behavior after bounce with respect to the structure of the PNS and its torus-like bulge is amplified in our ``Dl'' models. The hierarchy of the neutrino luminosities also mirrors the ranking of the central densities during the post-bounce evolution, with M1.42-J0 having the highest values and M1.61-J0.78-Dl and M1.91-J1.63-Dl the lowest ones, which are nearly identical, but model M1.91-J1.63-Dl develops a considerably more massive equatorial torus (see Table~\ref{tab:sims}).  

The luminosities in Figure~\ref{fig:lum-meanE} are the direction-integrated and energy-integrated energy fluxes (i.e., $\oint_{4\pi}\mathrm{d}\Omega\,r^2 F_{\nu_i}(r,\theta) = 2\pi \int_{-1}^{+1}\mathrm{d}\!\cos\theta \,r^2 F_{\nu_i}(r,\theta)$, measured in the lab frame at a radius of $r = 3500$\,km), i.e., the energy lost per unit of time by the PNS in the different neutrino species $\nu_i$. However, in the rotating models there is a large direction dependence of the isotropic-equivalent luminosities (i.e., the luminosities inferred by observers at different viewing angles, $4\pi\,r^2 F_{\nu_i}(r,\theta)$ at $r = 3500$\,km) of all neutrino species. This pole-to-equator variation develops quickly during the first 50--100\,ms post bounce, in our ``Dl'' models even faster, and continues to grow, reaching more than two orders of magnitude between pole (highest values) and equator (lowest values) a few seconds after bounce, and intermediate values at mid-latitudes that are a factor 2--5 higher than those at the equator. The ordering of the luminosities (for all neutrino species) of the different models relative to each other visible in Figure~\ref{fig:lum-meanE} is preserved for the isotropic-equivalent luminosities at the equator and at mid-latitude. However, as time advances, the isotropic-equivalent luminosities for all neutrino species near the pole in the rotating models exceed the (essentially direction independent) luminosities of the non-rotating model M1.42-J0. At 5\,s after bounce they are higher in all rotating models except in M1.61-J0.78-Dl and M1.91-J1.63-Dl, where we expect such a crossing to also occur at later times, when the luminosities in M1.42-J0 continue to decay while the cooling of the PNS further proceeds. 

The pole-to-equator contrast of the isotropic-equivalent luminosities increases with time, which goes hand in hand with a growing deformation of the (energy-averaged) neutrinospheres. The neutrinospheres (better: neutrino-``surfaces'') for the different neutrino species are defined by the locations where the effective optical depth for absorption and scattering reactions is equal to $\frac{2}{3}$, obtained by integrating the effective opacity $\kappa_\mathrm{eff} = \sqrt{\kappa_\mathrm{abs}(\kappa_\mathrm{abs}+\kappa_\mathrm{sc})}$, energy averaged over the energy-flux spectrum, along radial paths ($\kappa_\mathrm{abs}$ is the absorption opacity and $\kappa_\mathrm{sc}$ the opacity connected to scattering processes). While the neutrinospheres of all neutrino species are effectively spheres in the non-rotating model M1.42-J0 (apart from possible small-scale perturbations due to effects of convective flows), they have 3D shapes in the rotating models that resemble oblate spheroids or doughnuts with the high-density core of the PNS at their center. They extend far into the equatorial accretion belt down to densities between $10^{10}$\,g\,cm$^{-3}$ and $10^{11}$\,g\,cm$^{-3}$ for $\nu_e$ in the increasingly deleptonized, neutron-rich environment. The axis ratio between equator and pole becomes larger at later post-bounce times as the elliptical cross section of the $\bar\nu_e$-sphere becomes flatter, whereas the cross section of the $\nu_e$-sphere develops a more and more pronounced butterfly shape (see Figure~\ref{fig:M191-dens-neutrinospheres} for model M1.91-J1.09) or dumbbell shape (visible at several seconds after bounce in the AIC models of the low-density WDs, M1.42-J0.23-Dl, M1.61-J0.78-Dl, and M1.91-J1.63-Dl). Because of the highly aspherical structure of the rotating PNSs and the strong deformation of the neutrinospheres, there is a stark variation of the temperature and density along the neutrinospheres. Due to the steeper radial density profiles at the poles compared to the equatorial and mid-latitudinal directions (Figure~\ref{fig:M191-dens-neutrinospheres}), the neutrinospheric temperatures near the poles can be twice or even three times higher than those near the equator. Both this fact and the extreme oblate deformation of the neutrinospheres explain the large pole-to-equator differences of the isotropic-equivalent luminosities as well as directional variations of the mean energies of the radiated neutrinos that can reach a factor of 2 in some models and evolution phases with local maxima between 30$^\circ$ and 40$^\circ$ away from the rotation axis.

The emitted neutrino luminosities mainly scale with the maximum (and average) density and thus, due to compression heating, with the average temperature of the PNSs, whereas the mean energies of the radiated neutrinos are determined by the conditions at the neutrinospheres, which are highly aspherical in our rotating AIC models. Because of the complex shapes of the neutrinospheres of $\nu_e$ and $\bar\nu_e$, which change with time and imply substantial temperature variations across the surface, the behavior of the direction-averaged mean energies of the radiated $\nu_e$ and $\bar\nu_e$, again measured in the lab frame at 3500\,km, is much more complex than that of the luminosities. The radiated heavy-lepton neutrinos $\nu_x$ are mainly produced in the spheroidal high-density, high-temperature cores of the PNSs. Their mean energies reflect the rotation-dependent ordering of their luminosities, and their continuous decline signals the advancing cooling of the PNS core (Figure~\ref{fig:lum-meanE}, lower right panel). In contrast, this hierarchy applies for $\bar\nu_e$ only from $\sim$0.03\,s till $\sim$2.5\,s post bounce and for $\nu_e$ only approximately until $\sim$0.5\,s after bounce. At later times the formation of the massive equatorial accretion belts plays an increasingly more important role for the neutrinospheric conditions of these neutrino species. Since the mass of the torus-like belt varies between the models (see Table~\ref{tab:sims}) and its structure is determined by the angular momentum distribution in the collapsing WD, the detailed time evolution of the geometry and temperature distribution of the $\nu_e$- and $\bar\nu_e$-spheres is strongly model dependent.

Moreover, as the torus contracts and heats up, the mean energies of $\nu_e$ and $\bar\nu_e$ produced in the inner, high-density regions of the accretion belt can display an increase with time, which can be witnessed also for those neutrinos that escape from near the equator and from mid-latitudes. This behavior correlates roughly with the evolution periods at $t_\mathrm{pb}\gtrsim 3$\,s when episodes of enhanced inward flow of torus matter and associated eruptive mass ejection amplify the net mass-outflow rates as described in Section~\ref{sec:hydro-rap-rot-mod}. During these late-time phases the $\nu_e$ and $\bar\nu_e$ luminosities and mean energies exhibit short-timescale fluctuations due to accretion fluctuations of torus matter, the decline of $L_{\nu_e}$ and $L_{\bar\nu_e}$ may flatten (in particular in models M1.61-J0.78-Dl and M1.91-J1.63-Dl), and $\langle\epsilon_{\nu_e}\rangle$ and $\langle\epsilon_{\bar\nu_e}\rangle$ show slight tendencies of increase with unsteady longer-timescale variations. Therefore the rotation-dependent initial ordering of the direction-averaged $\langle\epsilon_{\nu_e}\rangle$ and $\langle\epsilon_{\bar\nu_e}\rangle$ between the different models is broken at later post-bounce times, and the species-dependent hierarchy $\langle\epsilon_{\nu_x}\rangle \gtrsim \langle\epsilon_{\bar\nu_e}\rangle > \langle\epsilon_{\nu_e}\rangle$, which holds for the non-rotating model M1.42-J0 at all times, is reversed to $\langle\epsilon_{\bar\nu_e}\rangle > \langle\epsilon_{\nu_e}\rangle > \langle\epsilon_{\nu_x}\rangle$ at some point in the rotating AIC models (Figure~\ref{fig:lum-meanE}). The post-bounce time when the energy crossing occurs is again strongly dependent on the angular momentum distribution in the initial WD, which governs the formation and evolution of the thick, equatorial accretion belt. For example, $\langle\epsilon_{\nu_x}\rangle \gtrsim \langle\epsilon_{\bar\nu_e}\rangle$ changes to $\langle\epsilon_{\nu_x}\rangle < \langle\epsilon_{\bar\nu_e}\rangle$ at $t_\mathrm{pb}\sim$1\,s in model M1.42-J0.23-Dl, at $t_\mathrm{pb}\sim$0.5\,s in M1.61-J0.47, and at $t_\mathrm{pb}\sim$0.2\,s in M1.91-J1.09, whereas $\langle\epsilon_{\nu_x}\rangle < \langle\epsilon_{\bar\nu_e}\rangle$ is fulfilled in models M1.61-J0.78-Dl and M1.91-J1.63-Dl during all of the plotted evolution. Similarly, $\langle\epsilon_{\nu_x}\rangle > \langle\epsilon_{\nu_e}\rangle$ transitions to $\langle\epsilon_{\nu_x}\rangle < \langle\epsilon_{\nu_e}\rangle$ at $t_\mathrm{pb}\sim$5\,s in model M1.42-J0.23-Dl, $\sim$3\,s in M1.61-J0.47, $\sim$2\,s in M1.91-J1.09, $\sim$1.5\,s in M1.61-J0.78-Dl, and $\sim$0.5\,s in M1.91-J1.63-Dl.

We close the overview of the neutrino properties of our models at this point and leave the presentation of more details for a dedicated, future paper. Finally, we note that \citet{Dessart+2006} also reported that higher neutrino luminosities and mean energies could be seen by observers located close to the polar directions compared to observers at viewing angles near the equator. However, the pole-to-equator variations obtained by \citet{Dessart+2006} in the faster rotating (1.92\,M$_\odot$) case of their two models were only about a factor of 2 for the luminosities and (10--30)\% for the radiated mean energies. Both of these polar-angle dependencies are much weaker than the corresponding effects described above for our rotating models. At post-bounce times similar to those reached by the calculations of \citet{Dessart+2006}, we see directional variations that can equal factors of 20--50 in the luminosities and up to $>$50\% in the mean energies (the exact values depend on the model and neutrino species), growing to more than a factor of 100 for the luminosities and up to a factor of 2 for the mean energies at later times. Moreover, the neutrinospheres exhibit shapes that are considerably different between the three considered neutrino species in all of our rotating models (Figure~\ref{fig:M191-dens-neutrinospheres} for model M1.91-J1.09 displays an example), in disagreement with the results shown by \citet{Dessart+2006} (see their figure~8 for their faster rotating model). 

The reasons for these substantial differences in the neutrino properties, despite largely similar hydrodynamic structures of our model M1.91-J1.09 and the 1.92\,M$_\odot$ simulation of \citet{Dessart+2006}, are unclear. They could be connected to the employed transport treatments, which are a two-moment M1 scheme in our \textsc{Alcar} code compared to multi-group flux-limited diffusion (MGFLD) without Doppler-velocity-dependent terms applied in the \textsc{Vulcan} code of the previous models. Also differences in the neutrino opacities might play a role in the calculations of the optical depths and neutrinosphere positions. It is well known that MGFLD leads to an increasingly more isotropic radiation flux for growing distances from the source. This phenomenon can be clearly noticed in figure~16 of \citet{Dessart+2006} already relatively close outside of the neutrinospheres at radii where the authors evaluated the neutrino emission (250\,km and 400\,km in their 1.46\,M$_\odot$ and 1.92\,M$_\odot$ models, respectively). More details on this point for neutrino transport with the \textsc{Vulcan} code can be found in \citet{Ott+2008}. On the other hand, due to nonlinear effects connected to the closure, M1 transport is known to suffer from the so-called beam-crossing problem or ``two-beam instability'' \citep{Olbrant+2012,Sadowski+2013,Skinner+2013}, where crossing light rays do not penetrate each other but are deflected to a collinear propagation direction. This can trigger a tendency to overestimate the polar flux in particular for geometries where the neutrino emission stems from a butterfly- or dumbbell-shaped neutrinospheric region \citep[see the appendix in][]{Just+2015a}. Although such numerical artifacts might mainly affect the $\nu_e$ flux geometry at late times in our simulations and less the $\bar\nu_e$ and $\nu_x$ transport, we find similar pole-to-equator differences for all three neutrino species. This disfavors the beam-crossing problem as a cause for the large variations with polar angle. Finally reliable information on the pole-to-equator contrast of the neutrino emission properties will require the solution of the full, angle-dependent Boltzmann transport problem in the highly aspherical AIC environment.

\begin{figure*}
    \centering
    \includegraphics[width=0.49\linewidth]{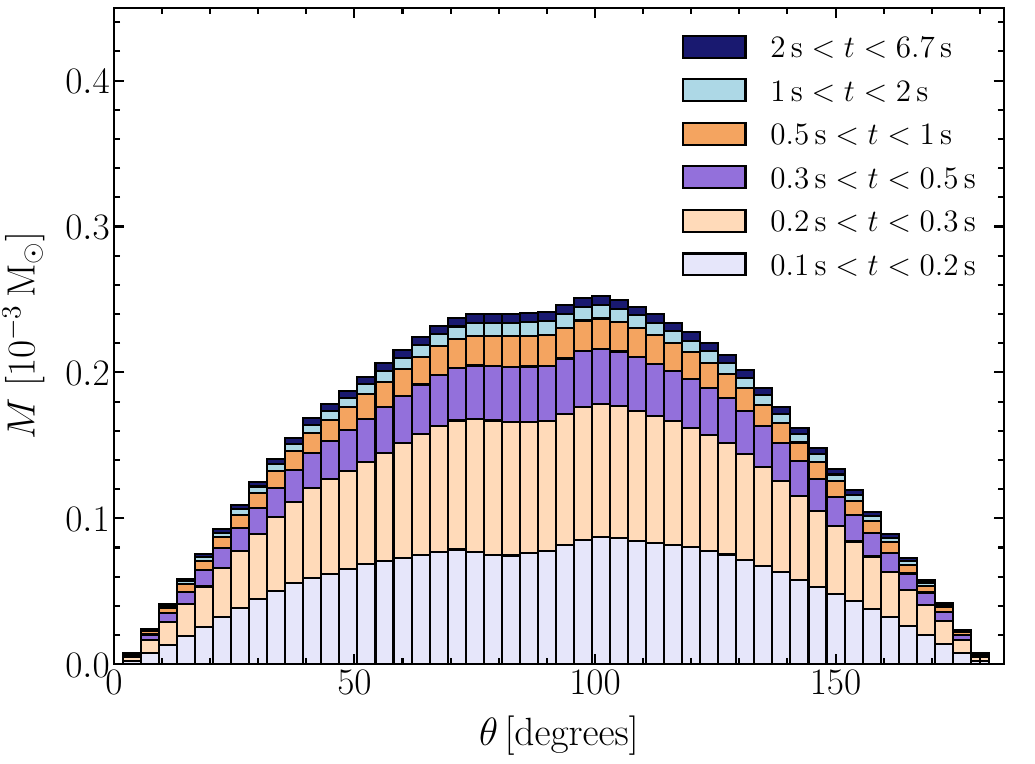}
    \includegraphics[width=0.49\linewidth]{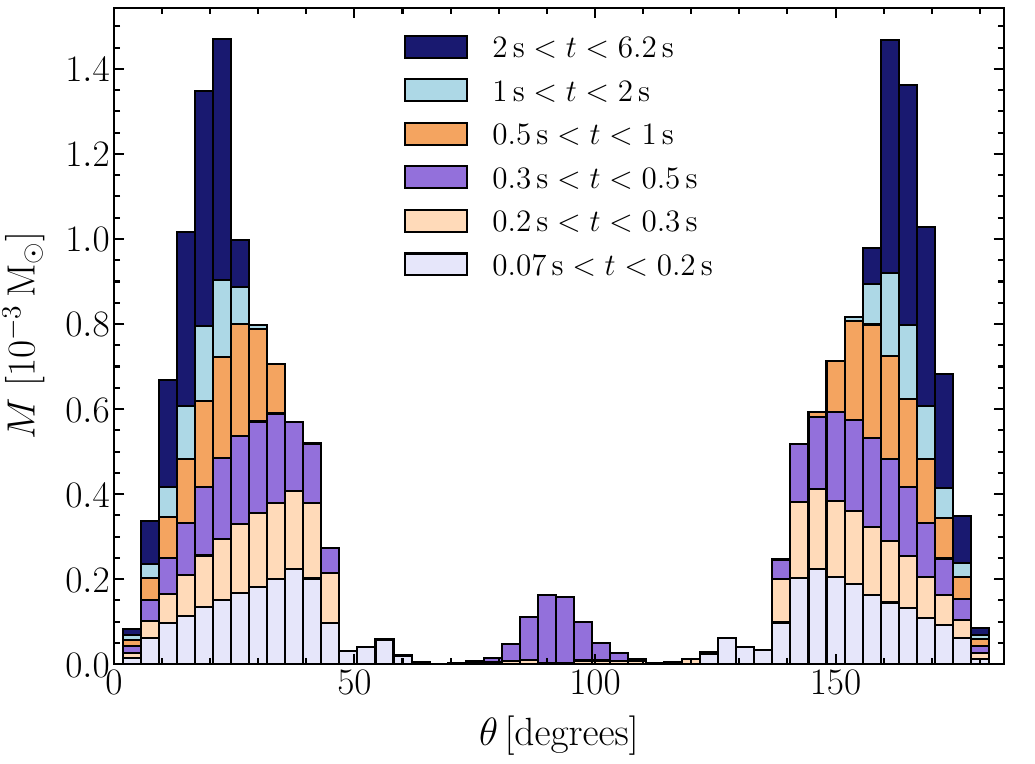}
    \caption{Ejecta mass versus polar angle $\theta$ for the non-rotating model M1.42-J0 ({\em left}) and the rapidly rotating model M1.91-J1.09 ({\em right}). The ejecta properties are determined for outflows through a sphere or spheroid placed just outside the initial non-rotating or rotating WD. The gas masses ejected in different time intervals after bounce are color-coded as indicated by the legends in both panels and stacked on top of each other so that each histogram bar displays the total mass expelled in the corresponding angular bin until the end of the simulation. The width of the bins is 3.75$\degree$ (i.e., twice the angular resolution). In the case of model M1.42-J0 one recognizes a nearly sinusoidal distribution ($\propto \sin\theta$) as expected for a non-rotating model with spherical outflow. The slight deviation from the ideal shape between $\sim$70$^\circ$ and $\sim$100$^\circ$ stems from a short initial phase of postshock convection with weakly anisotropic mass ejection. The predominantly polar outflows in model M1.91-J1.09 are confined to angles of less than 25$^\circ$--45$^\circ$ around the poles, decreasing with time because of the presence of the massive equatorial accretion torus and the abating power of neutrino heating, which drives the mass ejection. At larger distances, farther away from the spheroid of evaluation near the surface of the initial WD, the outflows spread around the accretion torus and the opening angles of the polar outflow cones widen.}
    \label{fig:M142+M191-theta-hist}
\end{figure*}

\begin{figure*}
    \centering
    \includegraphics[width=0.49\linewidth]{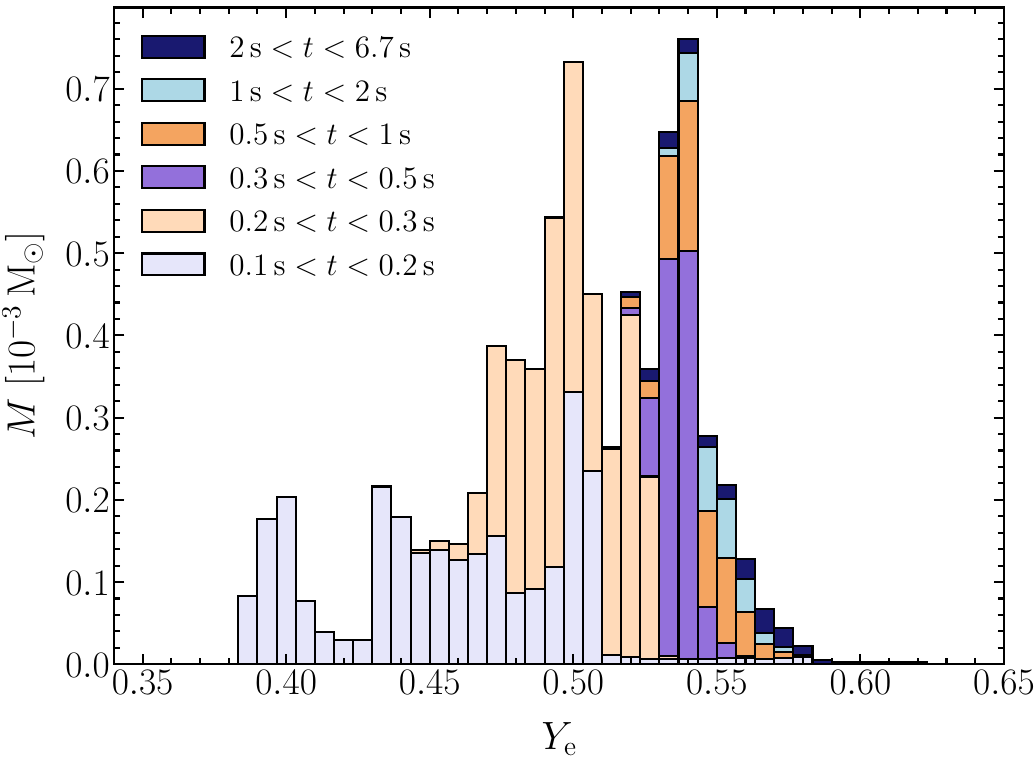}
    \includegraphics[width=0.49\linewidth]{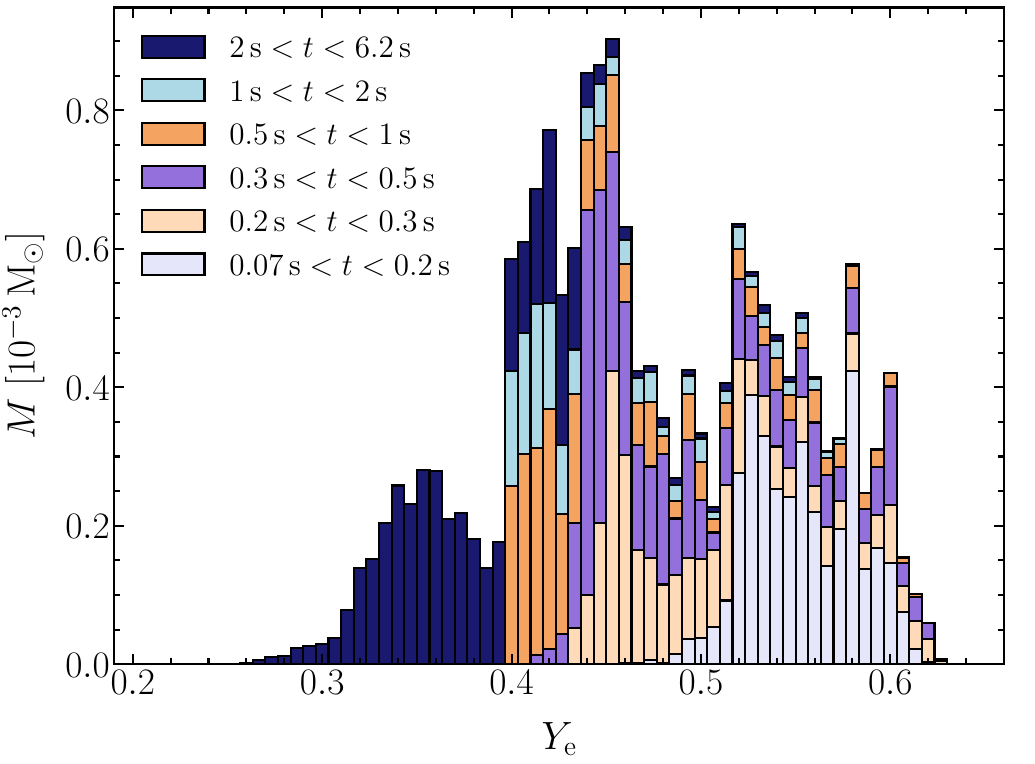}
    \caption{Ejecta mass versus electron fraction $Y_e$ for the non-rotating model M1.42-J0 ({\em left}) and the rapidly rotating model M1.91-J1.09 ({\em right}). The ejecta properties are determined for outflows through a sphere or spheroid placed just outside the initial non-rotating or rotating WD. The gas masses ejected in different time intervals after bounce are color-coded as indicated by the legends in both panels and stacked on top of each other so that each histogram bar displays the total mass expelled in the corresponding $Y_e$ bin until the end of the simulation. The width of the bins is $\delta \Ye \approx 0.0067$ (i.e., we use 15 bins per $\Delta Y_e = 0.1$ interval).}
    \label{fig:M142+M191-Ye-hist}
\end{figure*}

\begin{figure*}
    \centering
    \includegraphics[width=0.49\linewidth]{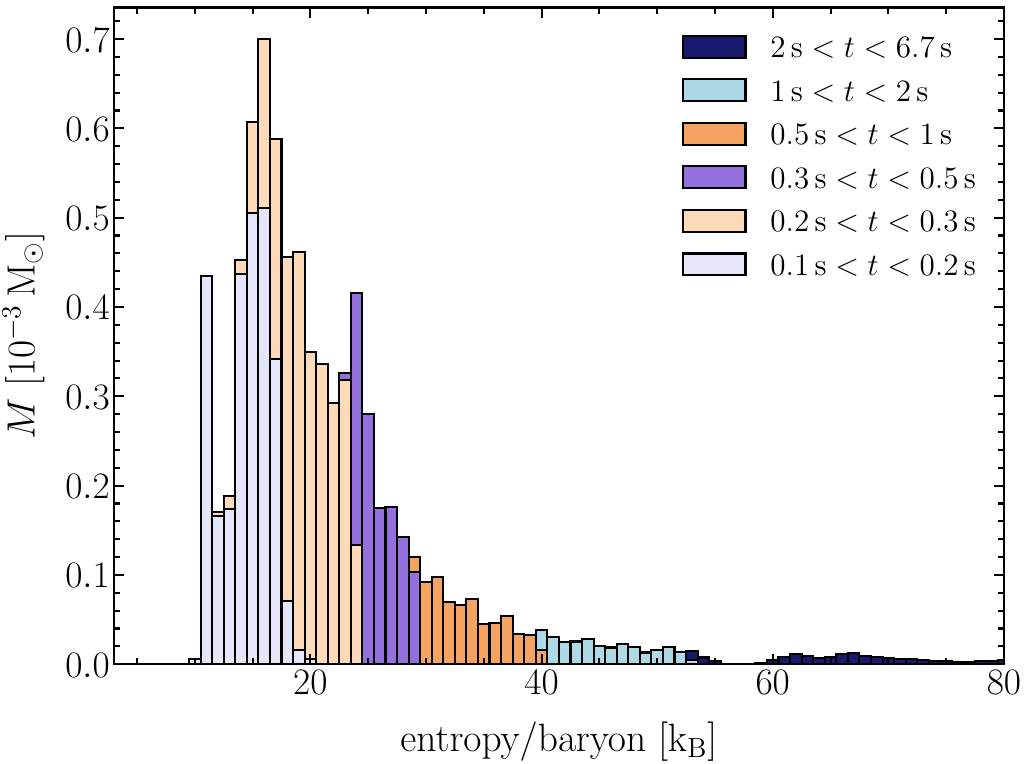}
    \includegraphics[width=0.49\linewidth]{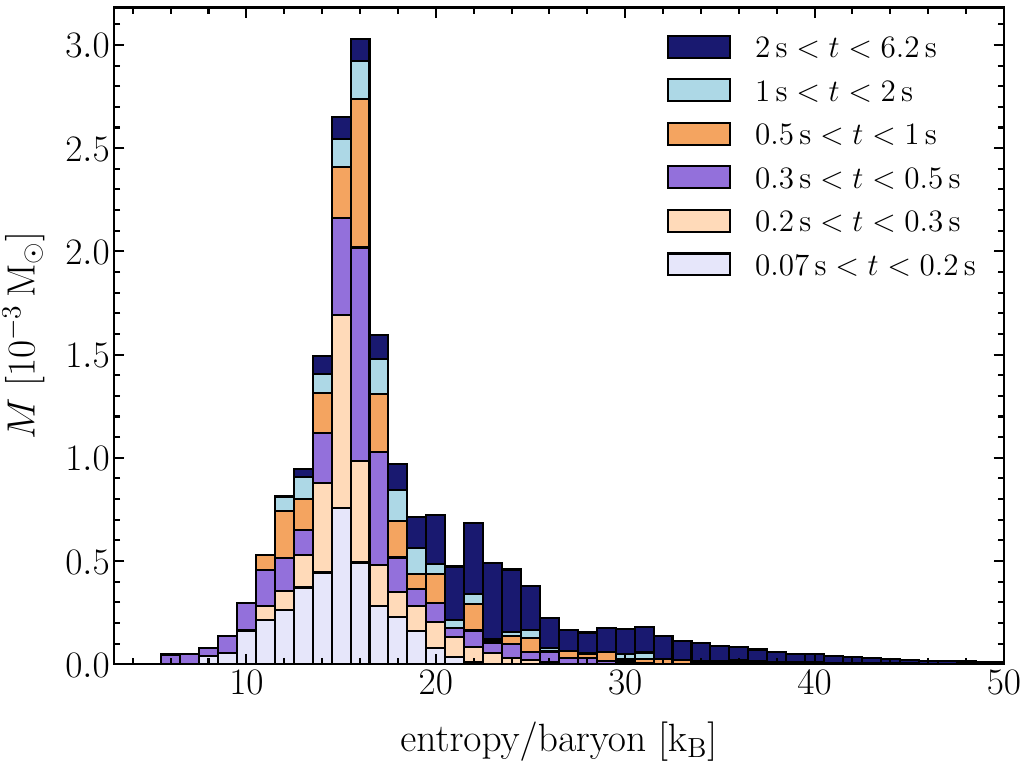}
    \caption{Ejecta mass versus entropy per baryon for the non-rotating model M1.42-J0 ({\em left}) and the rapidly rotating model M1.91-J1.09 ({\em right}). The ejecta properties are determined for outflows through a sphere or spheroid placed just outside the initial non-rotating or rotating WD. The gas masses ejected in different time intervals after bounce are color-coded as indicated by the legends in both panels and stacked on top of each other so that each histogram bar displays the total mass expelled in the corresponding entropy bin until the end of the simulation. The width of the bins is 1 in units of Boltzmann's constant $k_\mathrm{B}$.}
    \label{fig:M142+M191-entropy-hist}
\end{figure*}

\begin{figure*}
    \centering
    \includegraphics[width=0.49\linewidth]{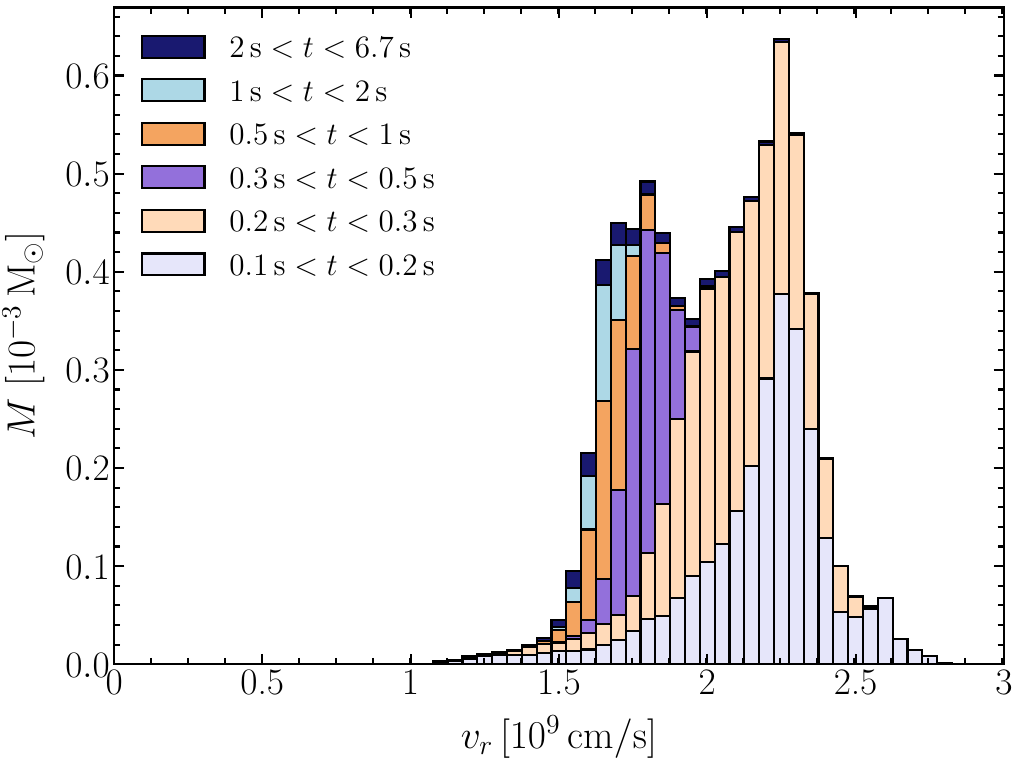}
    \includegraphics[width=0.49\linewidth]{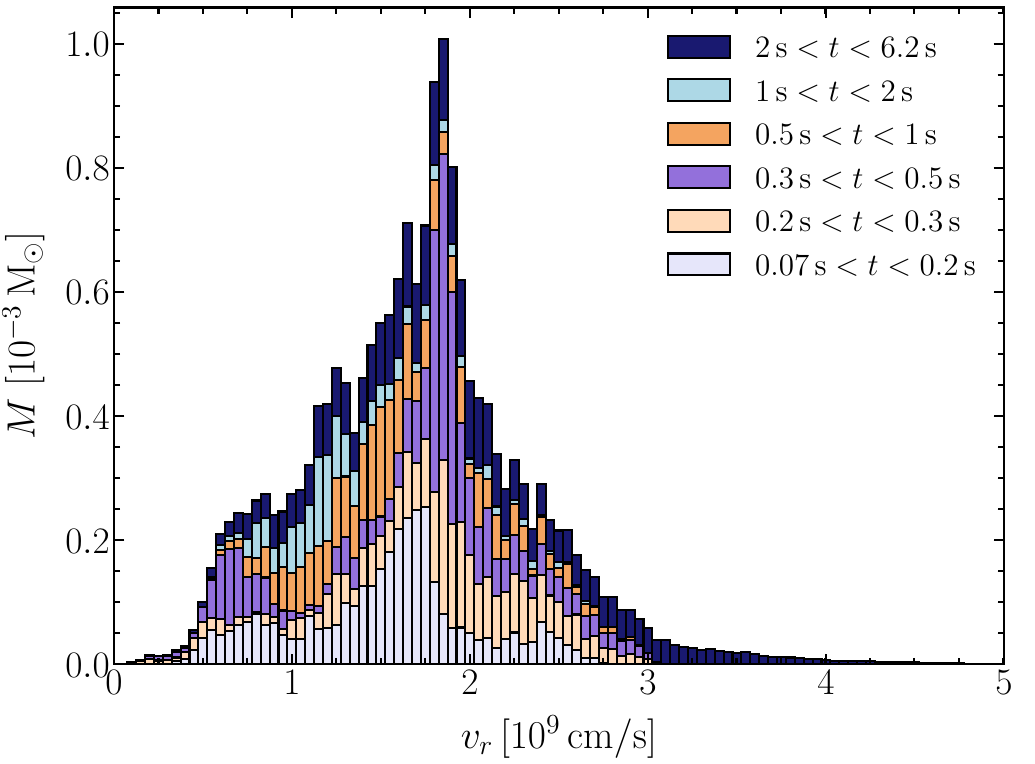}
    \caption{Ejecta mass versus radial velocity $v_r$ for the non-rotating model M1.42-J0 ({\em left}) and the rapidly rotating model M1.91-J1.09 ({\em right}). The ejecta properties are determined for outflows through a sphere or spheroid placed just outside the initial non-rotating or rotating WD. The gas masses ejected in different time intervals after bounce are color-coded as indicated by the legends in both panels and stacked on top of each other so that each histogram bar displays the total mass expelled in the corresponding entropy bin until the end of the simulation. The width of the bins is 0.5$\times 10^8$\,cm\,s$^{-1}$.\vspace{10pt}}
    \label{fig:M142+M191-velr-hist}
\end{figure*}

\begin{figure}
    \centering
    \includegraphics[width=\linewidth]{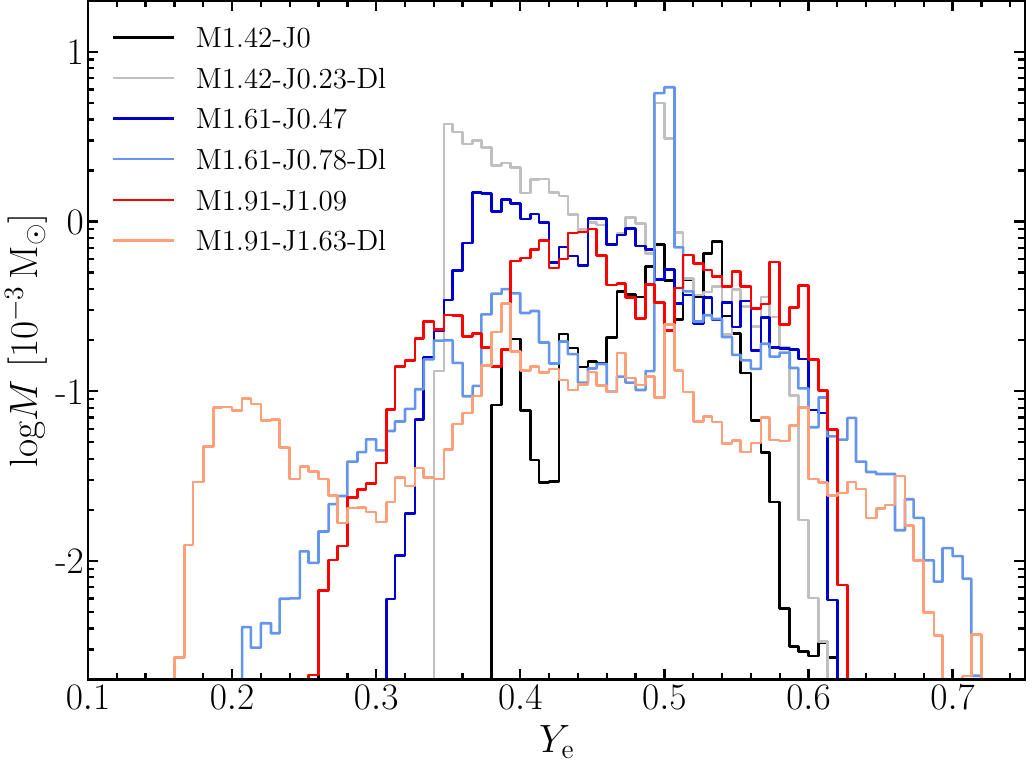}
    \caption{Ejecta mass versus $Y_e$ for all of our AIC simulations. The width of the bins is $\delta \Ye \approx 0.0067$ (i.e., we use 15 bins per $\Delta Y_e = 0.1$ interval). The ejecta properties are determined for outflows through a sphere or spheroid placed just outside the initial non-rotating or rotating WD. In contrast to Figure~\ref{fig:M142+M191-Ye-hist}, the mass scale is chosen to be logarithmic here.}
    \label{fig:Ye-hist-all}
\end{figure}

\begin{figure}
    \centering
    \includegraphics[width=\linewidth]{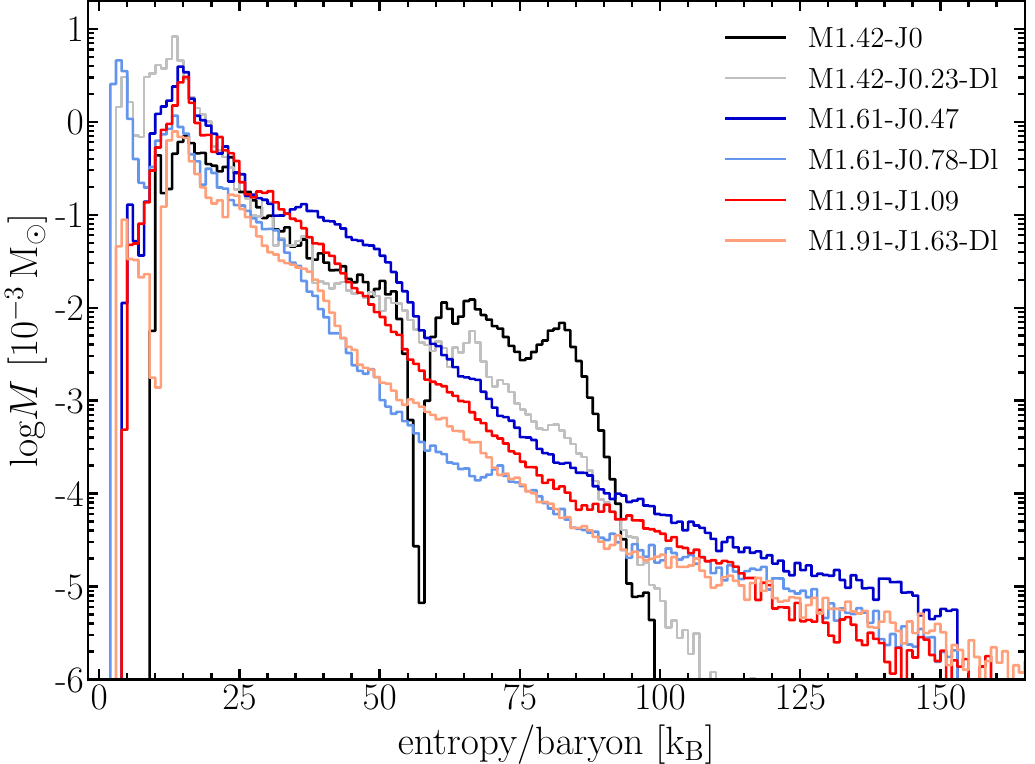}
    \caption{Ejecta mass histograms as functions of entropy per baryon in units of k$_\mathrm{B}$ for all of our AIC models. The width of the entropy bins is 1\,$k_\mathrm{B}$. The ejecta properties are determined for outflows through a sphere or spheroid placed just outside the initial non-rotating or rotating WD. In contrast to Figure~\ref{fig:M142+M191-entropy-hist}, the mass scale is chosen to be logarithmic here.}
    \label{fig:entropy-hist-all}
\end{figure}

\begin{figure}
    \centering
    \includegraphics[width=\linewidth]{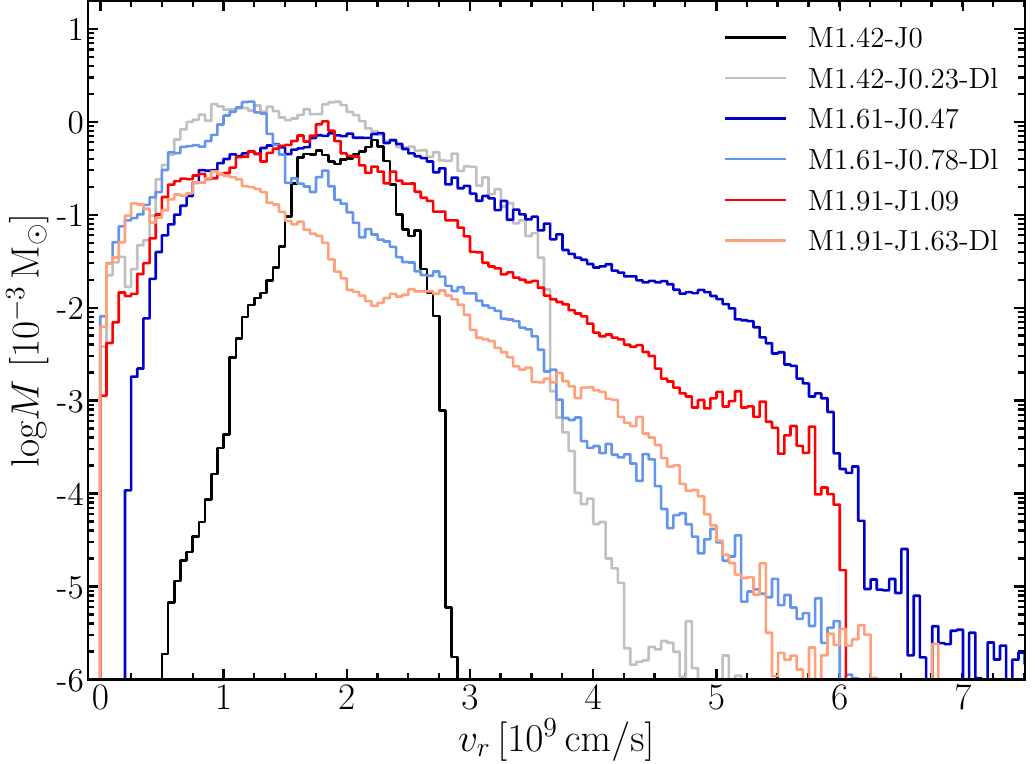}
    \caption{Ejecta mass histograms as functions of the radial velocity for all of our AIC models. The width of the velocity bins is 0.5$\times 10^8$\,cm\,s$^{-1}$. The ejecta properties are determined for outflows through a sphere or spheroid placed just outside the initial non-rotating or rotating WD. In contrast to Figure~\ref{fig:M142+M191-velr-hist}, the mass scale is chosen to be logarithmic here.}
    \label{fig:velr-hist-all}
\end{figure}
\begin{figure}
    \centering
    \includegraphics[width=\linewidth]{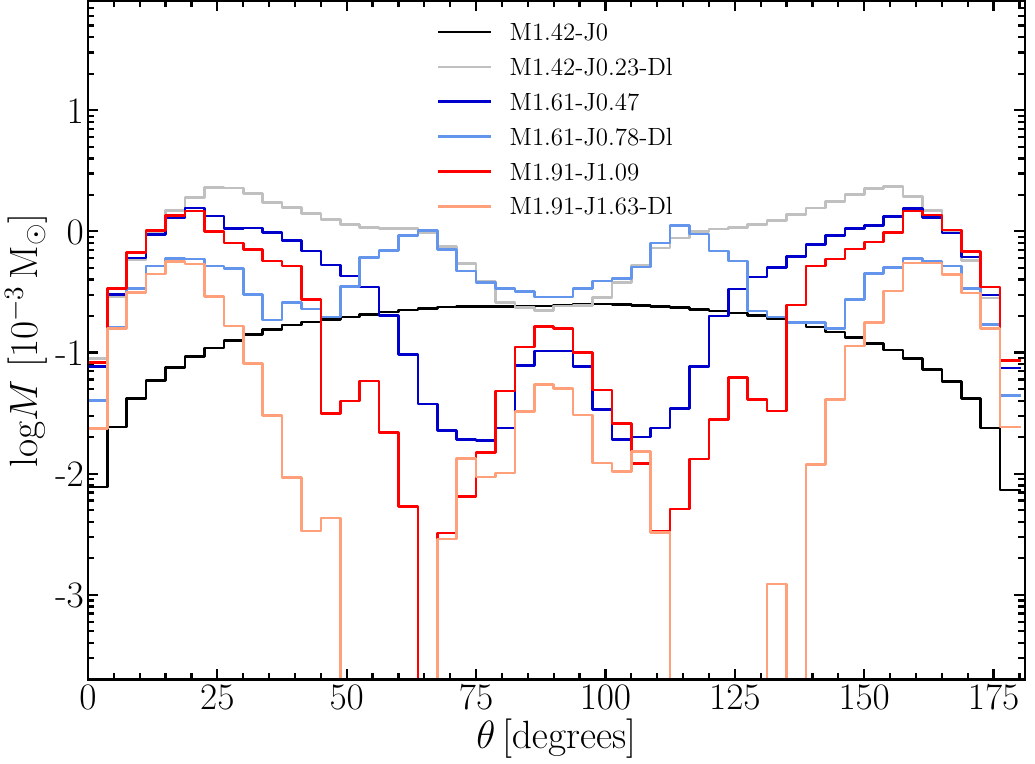}
    \caption{Ejecta mass histograms as functions of the polar angle $\theta$ for all of our AIC models. The width of the angular bins is 3.75$\degree$ (i.e., twice the angular resolution). In contrast to Figure~\ref{fig:M142+M191-theta-hist}, the mass scale is chosen to be logarithmic here. The ejecta properties are determined for outflows through a sphere or spheroid placed just outside the initial non-rotating or rotating WD. At larger distances the bipolar outflows spread laterally around the equatorial accretion tori and their opening angles become wider.}
    \label{fig:theta-hist-all}
\end{figure}

\subsection{Ejecta properties}

The nucleosynthesis-relevant properties of the ejecta are determined by the neutrino reactions in the infalling and outflowing matter, whereas our AIC simulations neglect possible effects of magnetic fields and viscosity. Table~\ref{tab:sims} lists the ejecta masses, which vary by nearly a factor of 10 at the end of our simulations and which are higher in the rotating models than in our non-rotating case. Model M1.91-J1.63-Dl is an exception in this respect, because it has extremely low neutrino luminosities, which render the neutrino-driven outflow particularly weak. We stress that our simulations take into account the energy release by nuclear recombination of free nucleons and $\alpha$-particles to heavy nuclei, which can help driving mass ejection \citep{Metzger+2009}. However, the ejecta masses in our AIC simulations are not the final values. Additional material will be lost from the equatorial, torus-like belts on secular timescales connected to viscous angular momentum transport due to MHD turbulence, in course of which a fraction of the torus matter will settle into the PNS whereas some other part will be expelled to carry away the excess angular momentum \citep{Metzger+2009}.  

The $Y_e$ in the outflow of our non-rotating model M1.42-J0 evolves similarly to that in the ejecta of ECSNe and ECSN-like events, where the neutrino-heated matter initially expelled in buoyant plumes is neutron-rich (see \citealt{Wanajo+2011, Wanajo+2018}, and \citealt{Stockinger+2020}). The subsequent, basically spherical neutrino-driven wind (Figures~\ref{fig:M142} and~\ref{fig:M142-dens-neutrinospheres} and Figure~\ref{fig:M142+M191-theta-hist}, left panel) becomes increasingly proton-rich at later times (see left panel of Figure~\ref{fig:M142+M191-Ye-hist}), accompanied by a continuous increase of the entropy per baryon in later wind ejecta (Figure~\ref{fig:M142+M191-entropy-hist}, left panel) and decreasing wind velocities at later times (Figure~\ref{fig:M142+M191-velr-hist}, left panel). The time evolution of the wind properties can be understood by their dependence on the evolution of PNS mass and radius and of the $\nu_e$ and $\bar\nu_e$ luminosities and mean energies \citep{Qian+1996}; it is much alike to that for a low-mass PNS computed with a mixing-length treatment of PNS convection in 1D \citep[resembling the results for an ECSN-like 9.6\,M$_\odot$ model with SFHo EoS displayed in figure~11 of][]{Mirizzi+2016}. However, the trend to $Y_e > 0.55$ is slightly more extreme in our simulations because of the overestimation of the mean $\nu_e$ energies in model M1.42-J0 mentioned in Section~\ref{sec:nuemission}. Figure~\ref{fig:M142-dens-neutrinospheres} shows that in the layers of the PNS exterior to the neutrinospheres, where the neutrino-driven wind forms and acquires its final conditions, the ratio of the number densities of $\bar\nu_e$ to $\nu_e$ ($n(\bar\nu_e)/n(\nu_e)$) increases at late post-bounce times. Therefore the trend to increasing wind-$Y_e$ is mainly facilitated by the growth of $\langle\epsilon_{\nu_e}\rangle$, which permits the dominance of absorption reactions $\nu_e + n \rightarrow p + e^-$ compared to $\bar\nu_e + p \rightarrow n + e^+$ in the outflowing matter. 

In summary, our non-rotating model M1.42-J0 yields ejecta with low electron fractions ($Y_e$ down to 0.38; Figures~\ref{fig:M142+M191-Ye-hist} and~\ref{fig:Ye-hist-all}), relatively low entropies ($s$ between 10 and 20\,$k_\mathrm{B}$ per nucleon; Figures~\ref{fig:M142+M191-entropy-hist} and~\ref{fig:entropy-hist-all}), and high velocities ($v_r$ up to nearly $3\times 10^9$\,cm\,s$^{-1}$; Figures~\ref{fig:M142+M191-velr-hist} and~\ref{fig:velr-hist-all}) from a short, initial phase of postshock convective overturn. The neutrino-driven wind that is ejected afterwards in this 2D simulation, which includes PNS convection, exhibits $Y_e$ values up to about 0.6, entropies up to about 100\,$k_\mathrm{B}$ per nucleon, and velocities between $5\times 10^8$\,cm\,s$^{-1}$ and $2\times 10^9$\,cm\,s$^{-1}$, in agreement with the neutrino-driven winds of low-mass PNSs from state-of-the-art 1D SN models that take into account PNS convection by a mixing-length treatment \citep{Mirizzi+2016,Pascal+2022}. We do not witness any effects of gravito-acoustic waves generated by PNS convection on the dynamics and nucleosynthesis conditions of the neutrino-driven wind. Therefore our long-term non-rotating model does not provide any support for the findings by \citet{Nevins+2024}, who recently investigated the consequences of convection-driven waves by a parameter study with steady-state wind solutions.

The deep, sharp dent at around 55\,$k_\mathrm{B}$ per nucleon in the entropy distribution of model M1.42-J0 in Figure~\ref{fig:entropy-hist-all} is caused by the reverse shock that forms due to the collision of the ejecta with the assumed CSM and their subsequent deceleration \citep[see][for a discussion in the SN context]{Arcones+2007}. This reverse shock travels backward in radius with time and passes the sphere where the outflow properties are monitored at a moment when the outflow has an entropy of nearly 55\,$k_\mathrm{B}$ per nucleon. Since the entropy of the flow through the reverse shock jumps by a few $k_\mathrm{B}$ per nucleon, the outflow moves on with an entropy close to 60\,$k_\mathrm{B}$ per nucleon, thus creating the dent between $\sim$55 and $\sim$60\,$k_\mathrm{B}$ per nucleon. This entropy depression is not filled by the later ejecta, because the entropy of the neutrino-driven wind in the non-rotating model increases with time and therefore all subsequent ejecta have higher entropies above the dent. The reverse shock occurring so early in the evolution is a consequence of the too high density of the CSM in our modeling setup and should therefore not be interpreted with respect to observations. While in the non-rotating model the reverse shock is a spherical feature that slowly and continuously progresses to smaller radii, the aspherical ejecta in the rotating models lead to less coherent reverse shock structures that are partly not perpendicular to the outflow and therefore have a reduced effect on the outflow properties \citep{Arcones+2011}. The slight deceleration and the small entropy increase ($\sim$5\,$k_\mathrm{B}$ per nucleon) of the quasi-spherical neutrino-driven wind in model M1.42-J0 are the only noticeable effects of the over-dense CSM on the results reported in our paper.

The $Y_e$ evolution with time is exactly opposite in the ejecta of all of our rotating models, where the earliest ejecta, expelled towards the polar directions (Figure~\ref{fig:M142+M191-theta-hist}, right panel), tend to be proton-rich and the later ejecta include increasingly bigger relative contributions of more and more neutron-rich matter (Figure~\ref{fig:M142+M191-Ye-hist}, right panel). On the basis of model M1.91-J1.09 as an example, Figures~\ref{fig:M191-J109} and~\ref{fig:M191-dens-neutrinospheres} show that proton excess occurs in a conical volume along the axis of the polar outflows. This proton-rich cone becomes more narrow with time and its proton excess decreases until $Y_e \ll 0.5$ holds in the entire outflow funnel at late times. The outflow regions with $Y_e < 0.5$ initially correlate with latitudes where the ratio of the $\bar\nu_e$ number density to the $\nu_e$ number density is highest (and also roughly with the directions where $\langle\epsilon_{\bar\nu_e}\rangle/\langle\epsilon_{\nu_e}\rangle$ is maximal), permitting relatively more frequent absorption reactions of $\bar\nu_e$ on protons in the ejected matter. The ratio of $n(\bar\nu_e)/n(\nu_e)$ increases with time in most models, consistent with increasingly more neutron-rich ejecta at later times. 

Therefore $\nu_e$ and $\bar\nu_e$ absorption on nucleons is favorable to keep $Y_e$ on the neutron-rich side at off-axis angles $\gtrsim$(10--20)$^\circ$ (model dependent and smaller at later times) and in the whole outflow funnels at very late times. Nevertheless, this cannot explain the growing neutron excess of the later ejecta, since the equilibrium $Y_e$ values due to the counter-acting $\nu_e$ and $\bar\nu_e$ absorption reactions are always around 0.5 or even higher. These equilibrium values are only reached in the narrow cone around the rotation axis, where the $\nu_e$ densities are highest and proton-rich or only mildly neutron-rich conditions hold most of the time (Figure~\ref{fig:M191-dens-neutrinospheres}). At larger angles the outflows possess much lower $Y_e$ than neutrino-absorption equilibrium would produce. This material is shed off the surface of the inner torus regions ($r \lesssim 100$\,km) at densities around $10^9$\,g\,cm$^{-3}$ and retains $Y_e$ values closer to those of its place of origin. Since this matter rotates very rapidly and is gravitationally loosely bound, it is easily ripped off by the interaction with the accelerating flow expelled directly from the vicinity of the PNS ($r \lesssim 30$\,km), and by weak neutrino heating, assisted by energy release through nucleon recombination. This material expands with velocities close to $10^9$\,cm\,s$^{-1}$ already at about 100\,km and thus avoids $Y_e$ to be lifted from its originally low values in the torus surface to values near the absorption equilibrium. In all of our rotating models the mass-ejection rates are considerably amplified at $t_\mathrm{pb} \gtrsim 3$\,s by eruptive mass-loss episodes connected to enhanced shear-flow activity in the boundary layers between torus and polar outflows (see Section~\ref{sec:hydro-rap-rot-mod}). During these late evolution stages the outflow-$Y_e$ drops well below 0.34, whereas all ejecta possess $Y_e \gtrsim 0.34$ in model M1.42-J0.23-Dl (Figure~\ref{fig:Ye-hist-all}), where the unsteady mass ejection phase sets in only shortly before the end of our simulation.

Models M1.61-J0.78-Dl and M1.91-J1.63-Dl exhibit the conspicuous feature that their $Y_e$ distributions extend up to more than 0.7, which is considerably higher than for the ejecta of any of the other models, where $Y_e$ reaches at most 0.62--0.63 (Figure~\ref{fig:Ye-hist-all}). This extremely proton-rich material in M1.61-J0.78-Dl and M1.91-J1.63-Dl is expelled with relatively low velocities (around $0.5\times 10^9$\,cm\,s$^{-1}$) in the first 0.1\,s after bounce. It crosses the outflow-evaluation surface with $v_r\sim (1-1.5)\times 10^9$\,cm\,s$^{-1}$ in the time intervals of $0.2\,\mathrm{s}<t_\mathrm{pb}< 0.3\,\mathrm{s}$ in M1.61-J0.78-Dl and $0.3\,\mathrm{s}<t_\mathrm{pb}< 0.5\,\mathrm{s}$ in M1.91-J1.63-Dl. The ejection of matter with such a high excess of protons happens in a phase when the newly formed, rapidly spinning PNS radiates mostly $\nu_e$. Since the ratio of $n(\bar\nu_e)/n(\nu_e)$ in the vicinity of the PNS is therefore extremely low, the absorption of $\nu_e$ on neutrons in the outflow dominates by far and pushes $Y_e$ to values up to 0.7 and beyond.

Model M1.91-J1.63-Dl is also exceptional in another respect, namely because it possesses a neutron-rich outflow in almost the entire polar funnels at $t\gtrsim2$\,s post bounce despite decreasing ratios of $n(\bar\nu_e)/n(\nu_e)$ in most of the funnel domain. In this model the neutrino luminosities and mean energies are extremely low at these late times, for which reason the ejecta simply retain $Y_e$ values close to their initial ones that they had when the outflow emerged from the neutron-rich accretion torus near the PNS surface. This model therefore exhibits the most pronounced low-$Y_e$ wing in its ejecta mass distribution compared to all of our other AIC simulations; its distribution reaches down to $Y_e\sim 0.15$, whereas the minimum values of $Y_e$ are about 0.2 for M1.61-J0.78-Dl, 0.25 for M1.91-J1.09, 0.3 for M1.61-J0.47, 0.34 for M1.42-J0.23-Dl, and 0.38 for the non-rotating case of M1.42-J0 at the end of our simulations (Figure~\ref{fig:Ye-hist-all}). 

Model M1.42-J0.23-Dl is in stark contrast to M1.91-J1.63-Dl by showing a proton-rich outflow in narrow axial funnels and neutron-rich ejecta only at larger angles away from the rotation axis until nearly the end of our simulation at $t_\mathrm{f}\sim 7.8$\,s. This difference can be understood by the relatively higher neutrino luminosities and mean energies in M1.42-J0.23-Dl. Overall, model M1.42-J0.23-Dl is closest to M1.42-J0 in the width of its $Y_e$ distribution (roughly covering the ranges [0.34,0.62] and [0.38,0.62], respectively), because the influence of rotation on its neutrino emission is weakest. The time when the entire outflow including the cone around the rotation axis becomes neutron-rich is therefore different between the models, and it happens sooner, the lower the neutrino luminosities and mean energies are.

For the AIC models of all low-density initial WDs one can witness prominent maxima in the mass-versus-$Y_e$ histograms around $Y_e\sim 0.5$ (Figure~\ref{fig:Ye-hist-all}). This neutron-proton-symmetric material originates from the extremely inflated and extended low-density torus around the PNS, and it is ejected mainly during the first 0.5\,s after bounce when the bounce shock and the neutrino-heated polar ejecta sweep around the torus and shed gravitationally loosely bound (because centrifugally supported) material off the torus surface. Since this matter has not fallen deep inward to reach higher densities and temperatures, electron captures were not fast enough to significantly reduce its $Y_e$ from the initial value of 0.5 that we assumed in our constructed WD models. This particular component of the ejecta is therefore directly affected by our choice of the initial WD conditions and will require closer investigation once self-consistently evolved AIC progenitors with more reliable $Y_e$ distributions become available.  

The highest entropies and outflow velocities are found close to the rotation axis, typically with a tendency of increasing values at later times (right panels of Figures~\ref{fig:M142+M191-entropy-hist} and~\ref{fig:M142+M191-velr-hist}). But the ejecta-mass distributions versus entropy per baryon and radial velocity are broad at all times, and the growing tails towards greater values are the only clear trend that can be witnessed for the rotating models as time advances. This contrasts with the systematic shifts of the distributions with time in model M1.42-J0, where the later ejecta add narrow segments of growing entropies and decreasing radial velocities to the histograms in the left panels of Figures~\ref{fig:M142+M191-entropy-hist} and~\ref{fig:M142+M191-velr-hist}, respectively. At most extreme conditions in the polar ejecta of the rotating models, the entropies reach up to more than 150\,$k_\mathrm{B}$ per nucleon, though for very little mass (Figure~\ref{fig:entropy-hist-all}), and the maximum expansion velocities are over 20\% of the speed of light (Figure~\ref{fig:velr-hist-all}). The mass distributions versus $Y_e$ and radial velocity of all rotating models are wider both on the low and the high sides than those of the non-rotating case of M1.42-J0. 

The distribution of the ejecta mass as a function of polar angle $\theta$ in this non-rotating model follows nearly perfectly the $\sin\theta$ behavior expected for spherically symmetric outflows (Figure~\ref{fig:M142+M191-theta-hist}, left panel). The short initial phase of postshock convection (Figure~\ref{fig:M142}, left panels) creates only a minuscule deviation (a small dip between $\sim$70$^\circ$ and $\sim$100$^\circ$ and a similarly weak hemispheric asymmetry) from the ideal shape due to slightly aspherical early ejecta (at post-bounce times $t_\mathrm{pb} < 0.3$\,s). This deviation is preserved at later times when the essentially spherically symmetric neutrino-driven wind adds more mass to the distribution. 

All rotating models exhibit strongly pole-dominated mass ejection (Figure~\ref{fig:M142+M191-theta-hist}, right panel, and Figure~\ref{fig:theta-hist-all}) when the outflow through a spheroid placed just outside the initial WD is monitored. The histograms display a deficit of mass near 0$^\circ$ and 180$^\circ$ with steeply rising slopes towards $\theta > 0^\circ$ and $\theta < 180^\circ$, respectively, because of vanishing solid angles and decreasing outflow densities near the rotation axis. The width of the polar outflow cones differs between the models. It is strongly influenced by the proximity of the torus to the origin of the mass ejection near the central PNS and by the extension of the accretion torus perpendicular to the equatorial plane. In this respect model M1.61-J0.78-Dl exhibits the peculiarity that the maxima of its ejecta-mass distribution are at large angles away from the pole, i.e., around 65$^\circ$ and 115$^\circ$. These maxima are connected to the early phase of the mass ejection in this AIC model of a low-density initial WD, in which the outgoing shock sweeps out matter along the surface of the extended accretion torus to form the prominent peak of the $Y_e$ distribution around $Y_e = 0.5$. A similar effect in model M1.42-J0.23-Dl leads to a bipolar outflow distribution with very broad and high maxima closer to the poles but no local maxima farther away from the rotation axis. In model M1.91-J1.63-Dl the bounce shock is too weak to strip off any larger amounts of the torus mass, for which reason the peak around $Y_e = 0.5$ in Figure~\ref{fig:Ye-hist-all} is low in this case and the mass distribution over $\theta$ displays a lack of ejecta at intermediate angles. 

The nonvanishing outflow contributions around the equator, i.e., the minima in the mass distributions of models M1.42-J0.23-Dl and M1.61-J0.78-Dl and small maxima around $\theta = 90^\circ$ of the other rotating models, are connected to small amounts of loosely bound matter at the outer edges of the tori that are expelled during the first few 100\,ms after bounce. This equatorial mass ejection is driven either by weak bounce shocks near the equator and/or when the shocks from the initial polar mass ejection have spread around the accretion torus and collide at the equator. Both effects push some original WD matter with $Y_e = 0.5$ outward near the equator. We repeat that at larger distances away from the PNS-torus remnant the polar outflows spread sideways and the outflow cones widen to encompass model-dependent half-opening angles between $\sim$45$^\circ$ and $\sim$85$^\circ$ and highest radial expansion velocities close to the poles, leading to substantial polar stretching and a prolate deformation of the ejecta clouds.

\subsection{Implications for nucleosynthesis}

A sizable part of the outflows of our AIC models consists of matter with $Y_e \gtrsim 0.485$, which can efficiently assemble $^{56}$Ni when the initially hot gas in NSE expands, cools, and undergoes a nuclear freeze-out. We therefore expect up to more than $10^{-2}$\,M$_\odot$ of this radioactive isotope being expelled in our models with the highest ejecta masses, with the exact amount strongly varying between the different models. Moreover, other stable and radioactive iron-group species (isotopes of Fe, Co, Ni, Zn) will also be formed for $Y_e \lesssim 0.485$ (for more detailed discussions, see, e.g., \citealt{Woosley+1992a,Hix+1999,Pruet+2003,Pruet+2004,Seitenzahl+2008,Metzger+2009}).

It has been known for long that neutron-rich ejecta with low $Y_e$ values, entropies per baryon, and outflow velocities as obtained in our rotating AIC models are potentially viable to produce heavy trans-iron nuclei through the rapid neutron-capture process (r-process). Early works have discussed nucleosynthesis under such conditions, for example, in the context of neutrino-driven winds of hot PNSs (see, e.g., Figure~10 in \citealt{Hoffman+1997}; also \citealt{Woosley+1992a,Meyer+1992,Woosley+1994,Witti+1994,Takahashi+1994}) and for the ejecta of NS mergers (see Figure~18 in \citealt{Ruffert+1997}; also \citealt{Freiburghaus+1999,Metzger+2010}). The latter figure shows that at typical entropies between several 10\,$k_\mathrm{B}$ and $\sim$100\,$k_\mathrm{B}$ per nucleon, elements up to mass numbers $A \sim 120$ including the first r-process abundance peak ($A$ around 80) can be copiously made for $Y_e$ as low as 0.3--0.35, where faster expansion of the ejecta and thus shorter expansion timescales shift this condition to slightly higher values. Nuclei with mass numbers up to $A \sim 150$ including the second r-process peak ($A$ around 120) require $Y_e \sim$0.2--0.25, the entire set of lanthanides ($139 \lesssim A \lesssim 175$) is strongly produced when $Y_e$ reaches down to 0.15--0.2, and the third r-process peak ($A$ around 190) including actinides is efficiently created when $Y_e$ is lower than this threshold. 

The $Y_e$ values in a smaller fraction of the ejecta in our non-rotating AIC model should therefore permit the formation of r-process nuclei up to mass numbers beyond 100, possibly marginally reaching silver \citep[see][]{Wanajo+2011}. In our rotating models we expect varying amounts of r-process nuclei with a strongly model dependent abundance distribution and a well developed third r-process peak only for model M1.91-J1.63-Dl and possibly marginally for M1.61-J0.78-Dl, involving only a minor part of the total ejecta. We leave a detailed evaluation of the nucleosynthesis in the outflows of our AIC models to future work, including an investigation of the question whether a solar-system like r-process abundance pattern can be obtained at least in a subset of the outflows (may be in some directions) of some of them.

\section{Summary, discussion, and conclusions}
\label{sec:summary}

We have presented the first long-term ($t_\mathrm{pb}\gtrsim 2$\,s) 2D neutrino-hydrodynamic simulations of PNS formation in AIC events of WDs, considering a set of six initial WD models with different masses, central densities, total angular momenta, and rotation profiles. Using the \textsc{Alcar} neutrino-hydrodynamics code with its elaborate, energy-dependent two-moment solver including fluid-velocity terms for the transport of $\nu_e$, $\bar\nu_e$, and $\nu_x$, our main goal was to determine the nucleosynthesis-relevant ejecta properties of AIC events, taking into account the effects of different degrees of rotation. In contrast to previous 2D and 3D simulations \citep[e.g.,][]{Dessart+2006,Dessart+2007,Abdikamalov+2010,LongoMicchi+2023}, our models do not follow the post-bounce evolution just for fractions of a second but cover much longer evolution periods up to times between $\sim$5\,s and $\sim$8\,s after bounce.

For our non-rotating AIC case we obtained an ejecta mass of nearly $8\times 10^{-3}$\,M$_\odot$ and a corresponding energy of $\sim$$10^{50}$\,erg, which is consistent with the neutrino-heated ejecta in state-of-the-art explosion models of ECSNe and ECSN-like events in the literature \citep[e.g.][]{Kitaura+2006,Janka+2008,Janka+2012,Melson+2015,Radice+2017,Stockinger+2020,Wang+2024}. Rotation in our other models has a strong influence on these properties, leading to ejecta energies between $\sim$$0.1\times 10^{50}$\,erg and $2.5\times 10^{50}$\,erg and ejecta masses ranging from slightly more than $5\times 10^{-3}$\,M$_\odot$ to over $5\times 10^{-2}$\,M$_\odot$ at the end of our simulations, depending on the initial conditions in the WD. We witnessed that higher ejecta masses roughly (but not strictly) correlate with higher energies, and the ejecta masses anti-correlate with the initial ratio $\beta_\mathrm{i}$ of rotational energy to gravitational energy. This can be understood by the facts that WD rotation facilitates mass ejection because of reduced gravitational binding of the WD matter, but very rapid rotation diminishes the radiated neutrino luminosities and mean energies and thus the neutrino heating that drives mass ejection.

The magnitudes of the neutrino luminosities (i.e., of the $4\pi$-integrated energy fluxes) and corresponding mean energies show an inverse correlation with the values of the ratio of rotational to gravitational energy of the inner core at bounce (and, roughly, also with this ratio for the gravitationally bound remnant at the end our our simulations). The lower $\beta_\mathrm{ic,b}$ and $\beta_\mathrm{rem,f}$ are, the more compact is the relic PNS with its equatorial accretion belt, and the hotter are the neutrinospheric regions for the different neutrino species on average. Therefore the rotating models also yield higher ejecta energies for lower values of $\beta_\mathrm{ic,b}$. Moreover, these models exhibit a huge pole-to-equator variation of the neutrino emission properties. For most extreme conditions the isotropic equivalent luminosities seen by observers at the poles can be more than two orders of magnitude higher than those observed from viewing angles near the equator at a few seconds after bounce, and the mean energies of the radiated neutrinos can vary by up to a factor of $\sim$2 with lowest values at the equator and off-axis maxima between 30$^\circ$ and 40$^\circ$. Interestingly, the species-dependent hierarchy of the emitted mean neutrino energies, which is $\langle\epsilon_{\nu_x}\rangle \gtrsim \langle\epsilon_{\bar\nu_e}\rangle > \langle\epsilon_{\nu_e}\rangle$ for the non-rotating model at all times \citep[for a detailed discussion in ECSN calculations with modern neutrino physics, see][]{huedepohl+2010,fischer+2010},\footnote{Note the subtle change in the order of $\langle\epsilon_{\bar\nu_e}\rangle$ and $\langle\epsilon_{\nu_x}\rangle$ at $\sim$0.5\,s after bounce when recoil energy transfer in neutrino-nucleon interactions is taken into account in the top panel of Figure~1 in \citet{huedepohl+2010} in contrast to the bottom panel of that figure and Figure~2 in \citet{fischer+2010}.} is reversed to $\langle\epsilon_{\bar\nu_e}\rangle > \langle\epsilon_{\nu_e}\rangle > \langle\epsilon_{\nu_x}\rangle$ at late times in our rotating AIC models. The post-bounce time when this occurs depends on the angular momentum distribution in the initial WD, which governs the formation and evolution of the thick accretion belt around the central, spheroidal, high-density PNS core.

A comparison of our AIC models with those of \citet{Dessart+2006} is most easily possible for their 1.92\,M$_\odot$ WD, which is quite similar to our initial WD of M1.91-J1.09. Our results show higher ejecta masses and energies as well as significantly different neutrino emission properties, e.g., higher pole-to-equator variations of the luminosities and mean energies. Moreover, in contrast to \citet{Dessart+2006} we witness shapes of the neutrinospheres that are very different between the different neutrino species. These discrepancies are probably linked to the different neutrino transport schemes (a two-moment M1 method in our case, flux-limited diffusion in the older models), although differences in the employed microphysics (neutrino opacities, nuclear EoS) might also have an influence. 

The ejecta properties depend crucially on the interaction of the escaping matter with the intense neutrino fluxes radiated by the hot PNS and its extended equatorial accretion belt. The non-rotating AIC model (similar to ECSN and ECSN-like explosions) possesses neutron-rich conditions in the earliest, neutrino-heated material that is thrown out in a convectively perturbed shell right behind the outgoing blast wave; the subsequent neutrino-driven wind ejecta become increasingly proton-rich as time advances. Significant rotation in the AIC models changes this result radically. In our rotating models the earliest ejecta in fast, wide-angle polar outflows are proton-rich because of high neutrino densities above the poles of the rotationally deformed AIC remnant. Only at later times, when the neutrino fluxes have declined, the ejecta become increasingly neutron-rich, retaining a neutron-to-proton ratio closer to the high initial values at the base of the mass ejection. The time evolution of the outflow-$Y_e$ in the rotating AIC models is therefore opposite to the general trend found in the non-rotating model. Since our simulations are the first ones with detailed, energy-dependent neutrino transport that followed the hydrodynamic post-bounce AIC evolution of rotating and non-rotating WDs for many seconds, this result has not been reported in any previous work (e.g., \citealt{Dessart+2006} could not address this point, because their models covered only a few 0.1\,s after bounce and the authors set 0.5 as an upper limit to $Y_e$ in the ejecta). 

The high neutron excess in a considerable fraction of the outflows with $Y_e$ values down to $\sim$0.15 suggests that our rotating AIC models can be sources of r-process elements, but the total mass and abundance distribution of the trans-iron nuclei will be strongly model dependent. This may permit useful constraints of the still uncertain AIC rate, if solar-system abundances of neutron-rich isotopes shall not be overproduced \citep{Woosley+1992,Fryer+1999}. Reliable conclusions, however, may be hampered by the severe sensitivity to the initial WD properties, which are currently not well determined. Similarly, depending on the initial WD conditions (electron fraction, rotation rate, angular momentum profile), a larger part of the expelled matter can have $Y_e \gtrsim 0.485$, shed from the outer layers of the massive accretion tori, and thus can end up in substantial amounts of $^{56}$Ni (possibly more than $10^{-2}$\,M$_\odot$ under favorable circumstances). The cloud of expanding ejecta will therefore be heated by the radioactive decay of $^{56}$Ni, other unstable iron-group isotopes, and r-process species, for which reason Ni-decay powered optical and infrared transients \citep{Metzger+2009,Metzger+2009a,Darbha+2010} and kilonova-like electromagnetic emission \citep{ChiKitCheong+2024} have been associated with AIC and MIC events. The true appearance is likely to be a combination of both possibilities and the luminosity evolution should vary significantly from case to case and with the viewing angle. A detailed investigation of these aspects will be the subject of future works. 

Better theoretical predictions of the elemental composition and electromagnetic signals connected to the AIC and MIC events are urgently needed to judge their potential for explaining unusual, kilonova-like transients such as AT2018kzr \citep{McBrien+2019,Gillanders+2020}, GRB\,211211A, and GRB\,230703A \citep{ChiKitCheong+2024}. Progress in this direction will also require better WD models at the pre-collapse stage from evolutionary calculations of AIC and MIC progenitor scenarios, because so far all hydrodynamical modeling of WD collapse has been started with initial conditions of mass, angular momentum and rotation profile, composition, and thermodynamic properties that are constructed on grounds of simple assumptions and partially imposed by hand. 

Interestingly, after several seconds of post-bounce evolution we estimate baryonic masses of only 1.163\,M$_\odot$, 1.308\,M$_\odot$, and 1.318\,M$_\odot$ for the PNSs in our models started with low-density initial WDs.\footnote{Note that the measurable gravitational masses are lower by the mass-equivalent of the energy carried away by neutrino emission. As shown, the neutrino losses depend on the rotation of the PNS-torus systems formed in the AICs. Since the compact remnants at the end of our simulations are not in their final states, we refrain from providing numbers for the gravitational masses.} In all three cases the PNSs are surrounded by massive accretion tori of 0.745\,M$_\odot$, 0.285\,M$_\odot$, and 0.056\,M$_\odot$, respectively, and major fractions of these tori are likely to be accreted onto the PNSs during the subsequent secular evolution when angular momentum is carried away by continuous stripping of sub-dominant fractions of the torus masses. We are therefore unable to provide reliable estimates for the gravitational masses of our final compact remnants, but consider it as unlikely that our AIC models could yield unusually small gravitational masses (around or less than 1.2\,M$_\odot$) similar to those inferred for the lowest-mass NSs in double NS systems \citep{Martinez+2015,Ridolfi+2019}, SN remnants \citep{Doroshenko+2022}, and low-mass X-ray binaries \citep{Oezel+2016}. However, magnetic fields enhance the mass loss \citep{Dessart+2007,ChiKitCheong+2024} and therefore WD collapse in AICs cannot be rigorously excluded as a channel for the formation of very-low-mass NSs. 

Naturally, our conclusions on the role that rapid rotation can play for NS birth, neutrino emission from hot PNSs, and the corresponding properties of the neutrino-driven (wind) ejecta are also relevant for collapsing stellar cores, if such cores can retain similar amounts of angular momentum as considered for our initial WDs. This, however, seems unlikely for single stars with magnetic fields \citep{Heger+2005} unless special circumstances apply such as very rapid rotation of the stars due to unusual birth conditions or binary interaction \citep{Woosley+2006} or due to low-metallicity, quasi-chemically homogeneous evolution \citep{Aguilera-Dena+2020}.

A caveat of our simulations is that they do not include the effects of magnetic fields or of any form of viscous angular momentum transport (beyond the consequences of a low level of numerical viscosity). Thus they should be considered as reference cases for comparison with future long-term simulations that will account for such additional physics, preferentially in 3D, to also allow for magnetically driven jets and torus outflows as well as triaxial (spiral and bar-mode) instabilities triggered by differential rotation. It seems unlikely that the consequences of magnetic fields will overrule our basic conclusions. This expectation receives support by the MHD calculations of \citet{Dessart+2007} and \citet{ChiKitCheong+2024}, although, for example, the anti-correlation between ejecta mass and initial rotation parameter $\beta_\mathrm{i}$ may be affected by magnetically-driven or spiral-arm-driven mass ejection in 3D models \citep[see][]{LongoMicchi+2023}. The strength of such effects is likely to depend on the angular-velocity profiles and magnetic fields in the pre-collapse WDs. However, a more detailed comparison of our results with those previous papers is difficult, because in addition to the MHD effects the simulations differ in many other aspects, e.g., in the initial WD setups (including, in particular, the initial angular-momentum distribution), the neutrino transport approximations and neutrino interaction rates, the EoS treatments, and the employed numerical grids and spatial resolution. All of these aspects can influence the outflow properties, for which reason our findings reveal major differences even relative to previous models without or with low magnetic fields. Meaningful conclusions on the influence of magnetic fields require a comparison of models with otherwise the same physics and numerics.

\vspace{5.0 truecm}\newpage
\section*{Acknowledgments}
We thank Robert Bollig for providing an extension of the EoS originally used in \textsc{Alcar}. EB would like to thank Andre Sieverding, Ulrich Steinwandel, Naveen Yadav, and Alkistis Zervou for useful discussions. The presented study was supported by the German Research Foundation (DFG) through the Collaborative Research Centre ``Neutrinos and Dark Matter in Astro- and Particle Physics (NDM),'' Grant No.\ SFB-1258-283604770, and under Germany's Excellence Strategy through the Cluster of Excellence ORIGINS EXC-2094-390783311. OJ acknowledges support by the European Research Council (ERC) under the European Union's Horizon 2020 research and innovation programme (ERC Advanced Grant KILONOVA No.\ 885281), by the DFG through Project-ID 279384907-SFB 1245 (subprojects B01, B06, B07), and by the State of Hesse within the Cluster Project ELEMENTS. HTJ would like to express special thanks to the Galileo Galilei Institute (GGI) for Theoretical Physics and the National Institute of Nuclear Physics (INFN) as well as the Mainz Institute for Theoretical Physics (MITP) of the Cluster of Excellence PRISMA+ (Project ID 390831469) for their hospitality and partial support during the final phase of this work.
The authors are grateful to the Leibniz Supercomputing Centre (LRZ; www.lrz.de) 
for computing time on the supercomputer SuperMUC-NG at LRZ under LRZ project ID pn49sa. Computer resources for this project were also provided by the Max Planck Computing and Data Facility (MPCDF) on the HPC systems Cobra, Draco, and Raven.

%

\vspace{5mm}

\software{\textsc{alcar} \citep{Just+2015,Just+2018,Obergaulinger2008},
          \textsc{Numpy} \citep{Jones2001},
          \textsc{Matplotlib} \citep{Perez2007}, 
          \textsc{IPython} \citep{Hunter2007} 
          }






\bibliography{ref}{}
\bibliographystyle{aasjournal}



\end{document}